\documentclass[a4paper,fleqn]{cas-sc}
\usepackage[utf8]{inputenc}
\usepackage[T1]{fontenc}
\usepackage{multicol} 
\usepackage{wrapfig}
\usepackage[numbers]{natbib}
\usepackage{siunitx}
\usepackage{caption}
\usepackage{subcaption}
\usepackage{url}

\def\tsc#1{\csdef{#1}{\textsc{\lowercase{#1}}\xspace}}
\tsc{WGM}
\tsc{QE}

\begin{document}
\let\WriteBookmarks\relax
\def\floatpagepagefraction{1}
\def\textpagefraction{.001}

\shorttitle{The Silicon Tracking System of the E16
experiment at J-PARC}


\title[mode = title]{The Silicon Tracking System of the E16 experiment at J-PARC: construction, installation and commissioning in beam test experiments}  

\author[a,b]{Dairon {Rodr\'iguez Garc\'es}}
\author[j]{Rento Yamada}

\author[c]{Kazuya Aoki}
\author[a,b]{Lady Maryann {Collazo S\'anchez}}
\author[a]{David Emschermann} 
\author[e]{Hideto En'yo}
\author[a]{J\"urgen Eschke} 
\author[a]{Ulrich Frankenfeld}
\author[a,b]{David {Guti\'errez Men\'endez}}
\author[a]{Johann M. Heuser}
\author[e]{Masaya Ichikawa}
\author[a]{Ralf Kapell}
\author[a]{Irakli Keshelashvili}
\author[a]{J\"org Lehnert}
\author[e]{Tomoki Murakami}
\author[d,e]{Shunnosuke Nagafusa}
\author[e]{Wataru Nakai}
\author[d,g]{Satomi Nakasuga}
\author[d]{Megumi Naruki}
\author[a]{Frederike Nickels}
\author[d]{Shuta Ochiai}
\author[c]{Kyoichiro Ozawa}
\author[a]{Dar\'io Alberto {Ram\'irez Zald\'ivar}}
\author[a]{Adrian {Rodr\'iguez Rodr\'iguez}}
\author[a,b]{Katia {Santos Marrero}}
\author[a]{Christian Joachim Schmidt}
\author[a,h]{Hans Rudolf Schmidt}
\author[a]{Kerstin Schuenemann}
\author[a,b,i]{Patryk Semeniuk}
\author[a,b]{Mehulkumar Shiroya}
\author[a]{Carmen Simons}
\author[a,b]{Sindhu Subramanya}
\author[a]{Oleksandr Suddia}
\author[f]{Tomonori Takahashi}
\author[a]{Maksym Teklishyn}
\author[a,b]{Alberica Toia}
\author[a]{Oleg Vasylyev}
\author[a]{Robert Visinka}
\author[j]{Yorito Yamaguchi}
\author[k]{Wojciech Zabolotny}

\address[a]{GSI Helmholtzzentrum f\"ur 
Schwerionenforschung GmbH,
  Planckstr. 1, 64291 Darmstadt, Germany}

\address[b]{Goethe University, Frankfurt am Main, Germany}

\address[c]{Institute of Particle and Nuclear Studies (IPNS), High Energy Accelerator Research Organization (KEK), Tsukuba, 305-0801, Japan}

\address[d]{Department of Physics, Kyoto University, Kyoto 606-8502, Japan}

\address[e]{RIKEN Nishina Center for Accelerator-Based Science, Saitama 351-0198, Japan}

\address[f]{Research Center for Nuclear Physics (RCNP), Osaka University, Ibaraki, 567-0047, Japan}

\address[g]{Advanced Science Research Center, Japan Atomic Energy Agency, Tokai 319-1195, Japan}

\address[h]{Physikalisches Institut, Eberhard Karls Universit\"at Universität T\"ubingen, Tübingen, Germany}

\address[i]{AGH University of Krakow, Krakow, Poland}

\address[j]{Physics Program, Hiroshima University, Higashi-Hiroshima 739-8526, Japan}

\address[k]{Institute of Electronic Systems, Warsaw University of Technology, Warsaw, Poland}

\begin{abstract}
The J-PARC E16 experiment aims to search for signatures of chiral symmetry restoration.
It studies in-medium modifications of vector mesons that decay via the dielectron channel.
The measurements use a high-intensity 30~\si{\giga\electronvolt} proton beam with C and Cu targets at rates up to 10~\si{\mega\hertz}. To achieve this, the experiment upgrades its tracking, by introducing innermost detector modules constructed with the same technology and procedures as the modules of the Silicon Tracking System (STS) of the Compressed Baryonic Matter (CBM) experiment at Facility for Antiproton and Ion Research (FAIR).

A total of 15 modules were assembled, tested, characterized and then installed in the E16 detector setup. The detector was commissioned in a beam test experiment at Tsukuba, where the detector modules could be exposed to a 3~\si{\giga\electronvolt} electron beam. In preparation for the beam test the modules were characterized and calibrated, and performance studies were accomplished to assess the quality of the setup. During beamtime, three modules were operated and illuminated in two planes by the electron beam.

This paper presents the results of the construction, characterization, commissioning, and operation of the E16-STS modules in beam test experiments.

\end{abstract}


\maketitle

\section{Introduction} \label{introduction}
\subsection{The E16 experiment at J-PARC}
The E16 experiment at the J-PARC Hadron Experimental Facility, Ibaraki, Japan \cite{Yokkaichi2022, Aoki2025E16Overview, Nakasuga2026E16First} studies dielectron spectra in 30~\si{\giga\electronvolt} proton-nucleus collisions to systematically examine the in-medium spectral changes of vector mesons in cold nuclear matter. The experiment aims to operate with a flux of 10$^{10}$ protons per 2-second spill at an interaction rate of 10~\si{\mega\hertz}.

The setup consists of a spectrometer with Silicon Strip Detectors (SSD) and a layer of 3 GEM Trackers (GTRs) \cite{Murakami2024GTR} for particle tracking in the magnetic field; and a Hadron Blind Cherenkov detector (HBD) and Lead Glass calorimeters (LG) for electron identification \cite{Nakasuga2026E16EID}. The detectors are arranged in sectors, each covering 30 degrees horizontally and vertically.

The E16 experiment aims to achieve 100 times more statistical data compared to the previous KEK-PS E325 experiment \cite{Naruki2006E325Omega, Muto2007E325Phi}. To reach this goal, the innermost tracking detector (SSD) is required to have a detection efficiency of 95\% over the entire momemtum coverage. Additionally, to achieve the mass resolution goal of approximately 5~\si{\mega\electronvolt}/c$^2$ a position resolution of 25~\si{\micro\meter} and time resolution of 6~\si{\nano\second} are required.

Due to unsatisfactory performance of the initially used SSD modules, which have large and thick frames, it was proposed to replace them with a system based on the technology developed for the Silicon Tracking System (STS) \cite{ststdr2013, RodriguezRodriguez2025, TEKLISHYN2025} of the Compressed Baryonic Matter (CBM) experiment at the Facility for Antiproton and Ion Research (FAIR) \cite{cbmphyprog2017}. This new system offers a lightweight configuration, high-rate capability, fine segmented design, and double-sided readout. The  specifications outlined in the Technical Design Report \cite{ststdr2013}, together with the results from the first prototype system \cite{Ramirez25} demonstrated that the CBM-STS modules achieve the position and timing resolution, as well as the hit rate demanded by the E16 specifications.

\subsection{The Silicon Tracking System module}
The STS module comprises a silicon microstrip sensor connected to custom-developed readout ASICs via a stack of microcables mounted on two front-end boards (FEBs). The double-sided, double-metal (DSDM) silicon microstrip sensor, produced by Hamamatsu Photonics K.K., Japan \cite{Hamamatsu}, is 320~\si{\micro\meter} thick and features 1024 strips per side with a pitch of 58~\si{\micro\meter}. The strips on the p-side are tilted by 7.5$^\circ$ with respect to those on the n-side, allowing two-dimensional position reconstruction from a single sensor.
The signals are routed from the sensor to the FEB by a stack of 32 low-mass aluminum–polyimide microcables. Each sensor side is read out by one FEB equipped with eight custom STS/MUCH-XYTER (SMX) ASICs \cite{KASINSKI2018225}, with each ASIC serving 128 strips, corresponding to the full set of 1024 strips per side. Each analog channel of the SMX ASIC has a dual path for energy and time measurements, featuring a 5-bit continuous flash ADC for signal amplitude and a timing discriminator with 3.125~\si{\nano\second} binning.

\subsection{The Silicon Tracking System setup}
\begin{figure}[h] 
        \includegraphics[width=\textwidth]{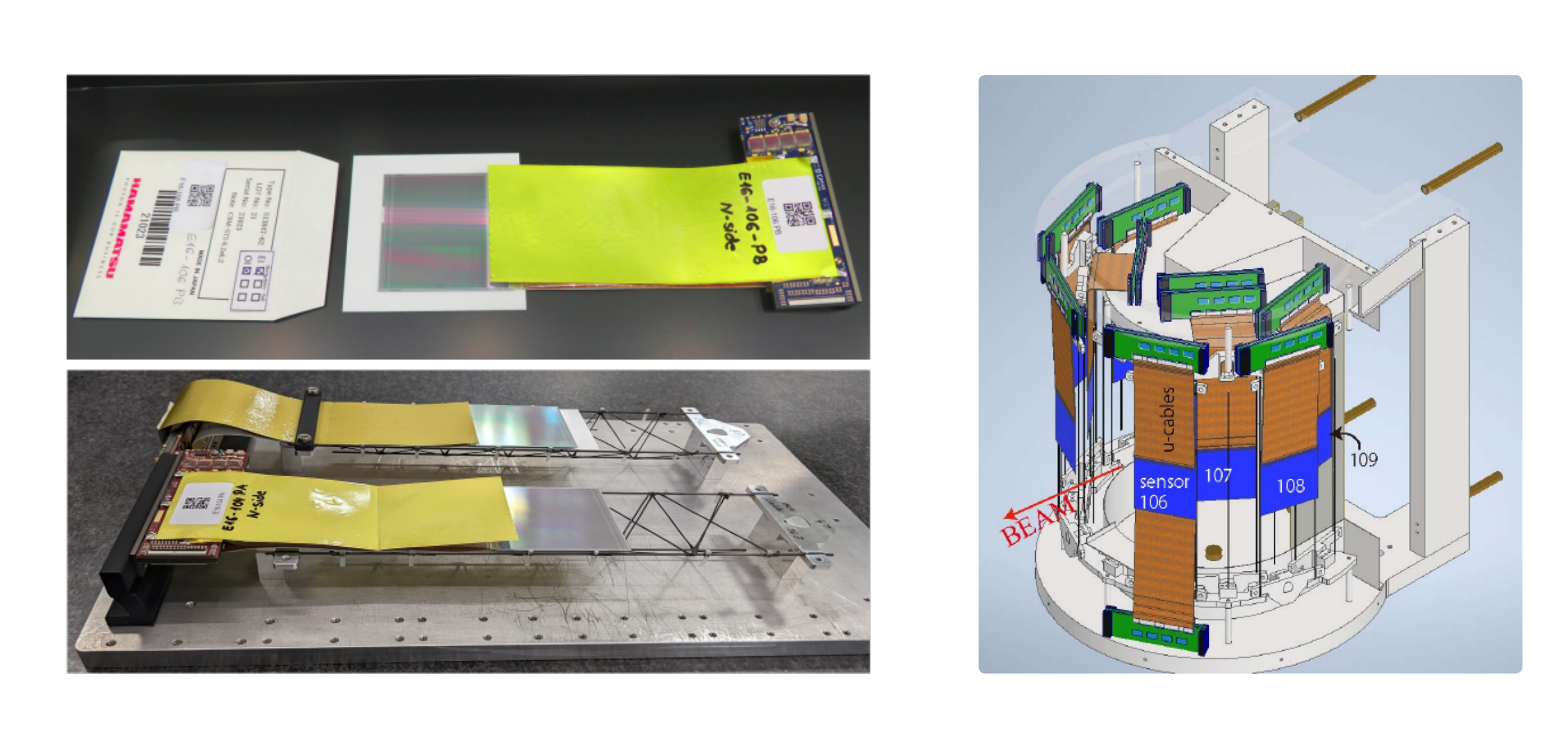}
    \caption{Left top: bare E16-STS module and bottom: fully assembled E16-STS modules mounted onto ladders and fixed onto a bottom plate ready for long-distance transportation. Right: a side view of the E16-STS chamber, several ladders with their modules are visible.}
    \label{fig:e16_setup}
\end{figure}

The E16-STS, a new setup based on CBM-STS modules, has been prepared to replace the SSD and provide the innermost space point in the E16 setup.
It consists of ten 62$~\times$~62~\si{\milli\meter}$^2$ sensors mounted on eight lightweight support structures based on carbon fibers, hereafter referred to as ladders.
Figure \ref{fig:e16_setup} summarizes the various integration stages of the E16-STS modules, including fully assembled modules mounted on ladders and a CAD rendering of their final integration into the E16-STS chamber. Eight sensors are positioned at the center of the ladders, while two ladders host one additional sensor. Each ladder, 231~\si{\milli\meter} tall, can hold up to three sensors for future upgrades. A target chamber containing experimental targets is installed at the center of the setup. The mechanical support structure is designed with minimal thick materials to avoid interference with the beam. The chamber's top and bottom plates are supported by two pillars and contain copper pipes for cooling the FEBs. To reduce light and electrical noise, aluminized-mylar sheets cover the inner and outer sides of the structure. Minor modifications were made to the original CBM-STS module design to accommodate the use of non-custom materials. 

\subsection{Integrating STS in a triggered system}
The J-PARC E16 experiment selects dielectron events from vector-meson decays in high-rate pA collisions. 
It uses discriminator signals from the GTR, HBD, and LG calorimeter, processed by trigger merging modules (TRG-MRG) and a Belle-2 UT3 decision board. The logic requires coincident hits in these detectors with a large opening angle of the dielectron pairs to suppress background. This reduces the raw interaction rate ($\sim$10~\si{\mega\hertz}) to a manageable trigger rate of about 1~\si{\kilo\hertz}.
However, the STS front-end ASIC operates in self-triggering mode, continuously sending all hit the information downstream. Although STS hits matching the E16 trigger could in principle be selected offline, this would produce an unmanageable data volume of about 300~\si{\tera\byte} per day. Therefore, a solution had to be implemented to couple the triggerless, free-streaming STS with the externally triggered system of the E16 setup.

This paper is organized as follows: Section 2 describes module production and characterization, including functional tests and response to a radioactive source; Section 3 covers the commissioning and operation of the E16-STS chamber in beam-test experiments at the KEK PF-AR beamline; Section 4 presents the results, including signal amplitude, time resolution, and position resolution; finally, Section 5 discusses the strategy to couple the STS free-streaming system with the E16 triggered setup.

\section{Module production and characterization}
The module production was carried out at GSI Detector Laboratory, with 15 pre-series modules of the CBM-STS. The assembly and components tests followed the final STS series assembly protocol described in reference \cite{Irakli}. 
Due to geometric constraints, the modules were assembled using 62~$\times$~62~\si{\milli\meter}$^2$ silicon sensors and microcables ranging from 12 to 20~\si{\centi\meter} in length \cite{LTU}. The FEB used is the third version (FEB-8v3), which is part of the pre-series modules leading up to the series production of the STS. The v3 version introduces unique features, including a new data cable connector and an onboard high voltage (HV) filtering system (2-stage RC filter). Additionally, the signal return path between the p- and n-side boards was implemented using a multi-wire cable.
During the assembly process all the module components, such as the sensor, the SMX ASIC and the microcables bonded to the ASIC by TAB bonding technology, were tested.

\subsection{Functional characterization}
After being assembled, each module undergoes a detailed and thorough characterization following a testing 
protocol based on the approach detailed in \cite{RodriguezRodriguez2024}. An initial visual inspection serves to exclude any damage like e.g.
sensor scratches or missing FEB components. 
For the functional characterization, the module is mounted on a carrier that
provides mechanical support, protection, and the necessary power and data connections. The carrier is then placed in a light-tight aluminum box for performance evaluation and calibration. The grounded box also offers protection against
electromagnetic interference. The use of the FEB8-2v3, which includes HV filtering and a signal return path, required adjustments to the ground and power scheme of the setup. These modifications were incorporated into
the test box to accommodate the E16 modules. In addition, the test box provides all essential service interfaces, including
low voltage and sensor bias voltage, readout, and cooling for the front-end electronics. The FEBs are connected to a CERN-GBT based common readout board (C-ROB) via flexible flat cables, with data aggregated and transmitted through a high-speed optical link to the data processing board, implemented on an FPGA-based (AFCK) board. A Python-based code manages the full readout functionality,
including synchronization, configuration, calibration, and noise measurements.
Each module is powered on, and the output voltages and current consumption for each voltage regulator on the FEBs are measured to exclude electrical shorts or malfunctioning components on the FEBs.
The evaluation of the modules performance consists of different measurements, which are presented below and detailed in separated subsections.

\subsubsection{Measurement of the current-voltage characteristics}
\begin{figure}[h] 
    \centering
    \includegraphics[width=0.85\textwidth]{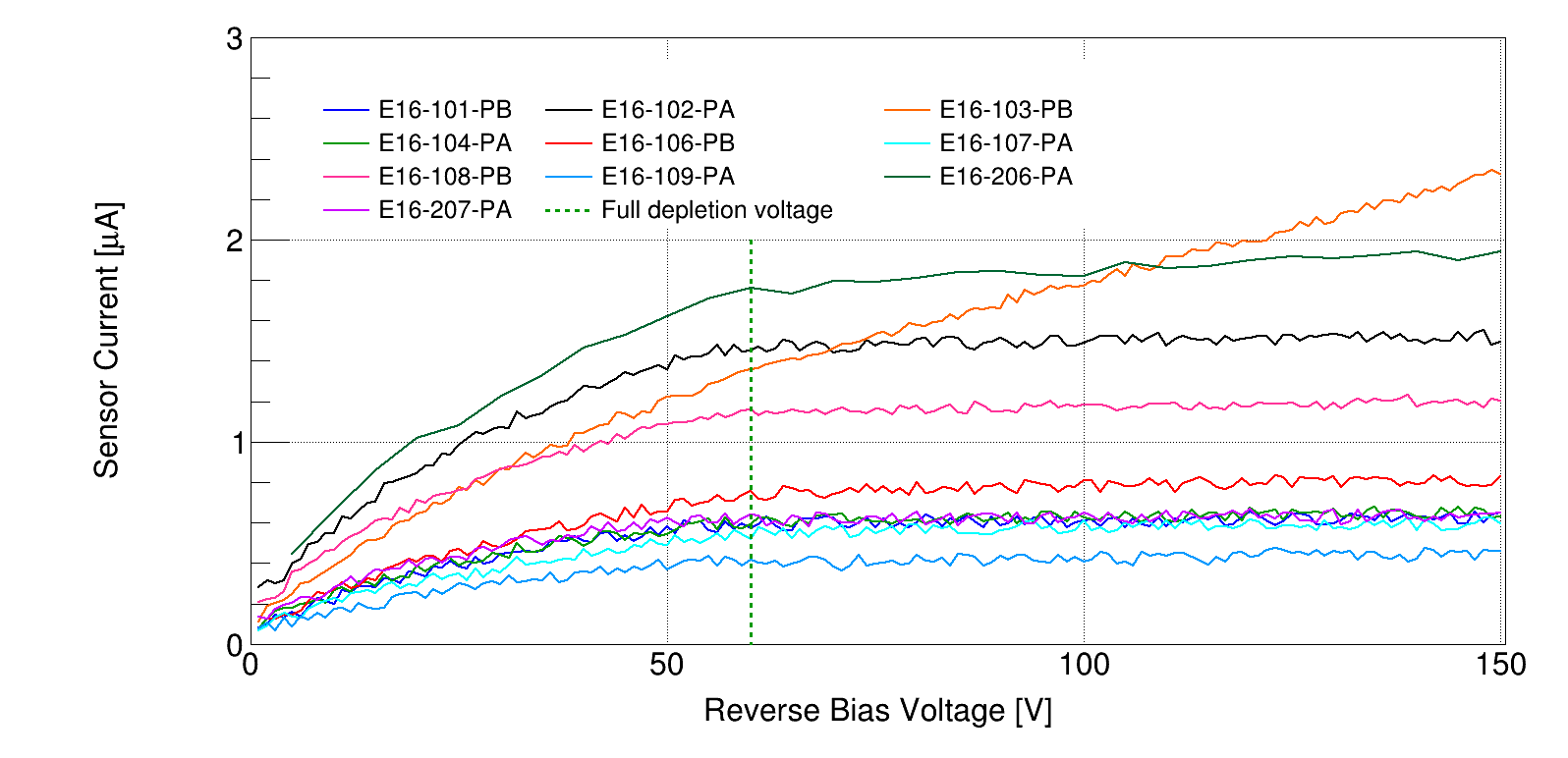}
    \caption{Current-voltage sensor measurement for the first ten E16-STS modules. The full depletion voltage at $60~\si{\volt}$ is shown with a vertical dashed line.}
    \label{fig:E16_IV}
\end{figure}

One of the most important tests is the sensor-reverse bias voltage scan (IV), 
which measures the sensor leakage current (I) as a function of applied voltage (V). This step provides a measurement of the current-voltage characteristic of the double-sided silicon sensor after assembly with all the other module components, and it allows to identify full depletion voltage, and early current-breakdowns, the latter may prevent the operation of the module. 
Figure~\ref{fig:E16_IV} shows the results of the IV scan. The depletion voltage is reachable around 60~\si{\volt} in all cases, consistent with bare sensor measurements and the manufacturer's data. After the depletion voltage a long operational plateau is visible, where the current remains stable and low (below 3~\si{\micro\ampere}) up to the operational value of 150~\si{\volt}. 
Different levels of leakage current between sensor of the same size are measured, because the IV results have not been corrected by temperature.

A significant difference is observed for module E16-103-PB, which shows a linear behavior beyond the full depletion voltage. This is a consequence of the configuration used for measuring the IV curve for that specific module, and can be explained as follows: as the low voltage (LV) power supply was connected and it is not entirely floating, its protection circuit provides a resistive leakage path to the biasing circuit, resulting in a linear behavior proportional to the resistance of the protection circuit.       

\subsubsection{Characterization of the E16-STS front-end electronics}
The data link initialization and synchronization allow verifying the communication with all data links in each ASIC (two per in this case) on the module, ensuring proper connection and reliable operation.
At the same time, the unique e-fused identification number is read out.

\begin{figure}[h!]
    \centering
    \begin{multicols}{2}
        \includegraphics[width=\linewidth]{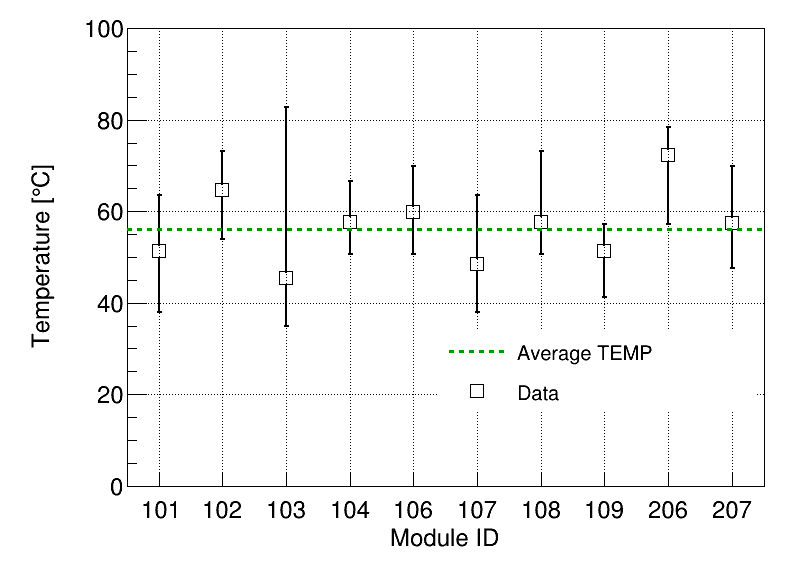} \par
        \includegraphics[width=\linewidth]{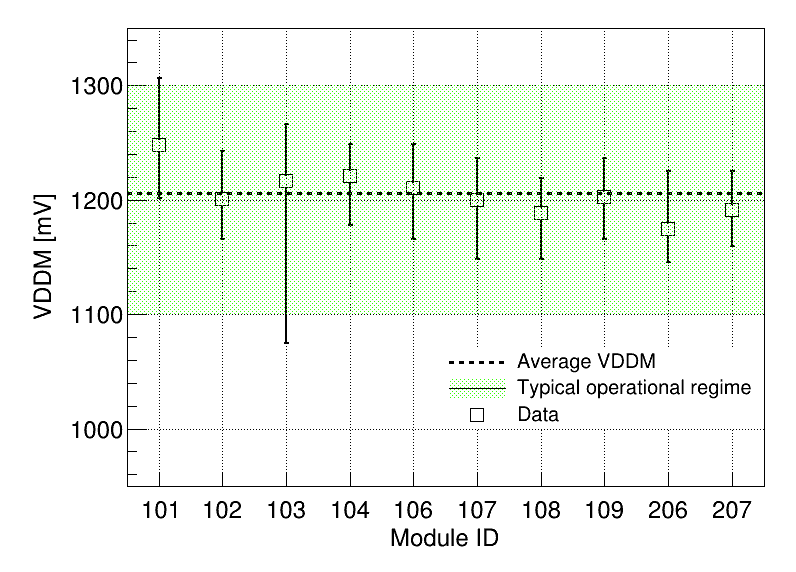} \par
    \end{multicols}
    \caption{Left: Temperature and Right: potentials for each module measured with the built-in diagnostic circuit. Every point is the average among 16 SMX ASICs and error bars represent the minimum and maximum measured values.}
    \label{fig:temp_vddm}
\end{figure}

Multiple write/readback actions ensure proper
settings of the register configuration parameters.
The power consumption, temperature, and potential values are verified. 
Measurements are performed for each of the 16 ASICs in every module using the built-in diagnostic circuit with average transfer functions for voltage and temperature. This explain the fluctuations within a module. The observed temperature variations between modules can be explained by differences in the E16-STS module form factors. In particular, some modules have equal microcable lengths for n- and p-side, which had implications on the way the FEB is cooled in our testing setup, and therefore expressed in module-to-module temperature fluctuations. The measured values are shown in Figure \ref{fig:temp_vddm} and in line with the expected ones. 

\subsubsection{Calibration of the ASICs analog front-end}
The calibration of the Analog Front End (AFE) circuits is performed to adjust individually the threshold of each channel's
discriminators (31 ADC + 1 TIME), using a combination of global ADC and FAST reference potentials, a resistor chain that generates individual thresholds, and trims DACs to fine adjustments. 

The calibration ensures a linear and homogeneous response across all channels. Without proper calibration, channel-to-channel variations in gain or discriminator threshold would introduce systematic distortions in the reconstructed charge and timing information. The procedure is performed by adjusting trim DACs to find the middle point (mean value of the so-called S-curve, representing the effective threshold) in the discriminator response at a fixed injected charge. The gain is determined by fitting the threshold values to the injected charge, which also verifies the linearity of the ADC after calibration \cite{RodriguezRodriguez2024}.

The calibration of the analog measuring circuits is performed with fully depleted sensors biased at 150~\si{\volt}. 
The LV was provided by the WIENER MPOD OMPV.8016 \cite{wiener}, an 8-channel module designed for reliable low-voltage operation, while the HV was supplied by an ISEG device version EHS 203 05x-K01 Customized \cite{iseg}, the same foreseen for the operation of the CBM-STS detector. The FEB temperature was maintained under control with a simple, but efficient cooling system composed of three basic elements: a cooling block, connecting pipes, and a LAUDA chiller \cite{lauda}. 

\begin{figure}[h!]
    \centering
    \begin{multicols}{2}
        \includegraphics[width=\linewidth]{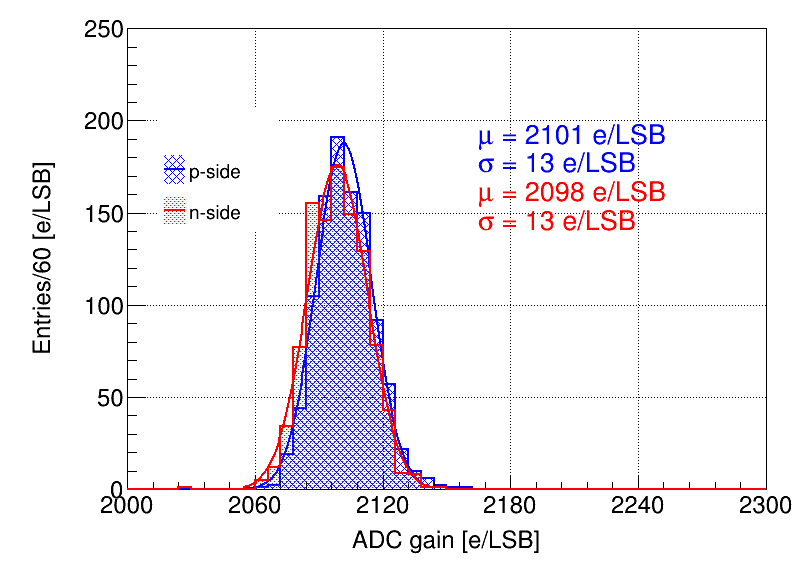} \par
        \includegraphics[width=\linewidth]{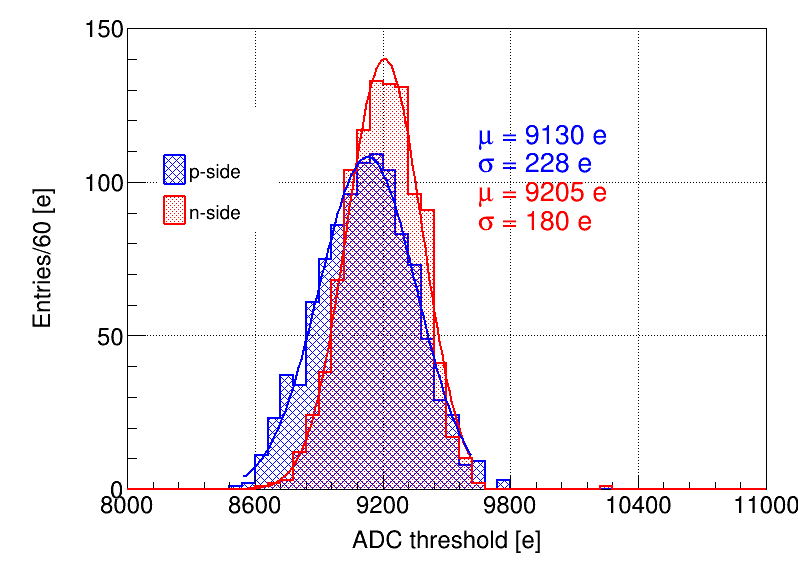} \par
    \end{multicols}
    \caption{Left: ADC gain and Right: threshold distribution for all channels across module. Mean and sigma values determined for each side through a Gaussian fit.}
    \label{fig:adc_gain_thr}
\end{figure}

Figure~\ref{fig:adc_gain_thr} shows for an exemplary E16-STS module (E16-106-PB) the distribution after calibration of the ADC gain and thresholds among all the channels for the 16 ASICs. The average gain values for each side corresponds to 2098~e/LSB and 2101~e/LSB, for n- and p-side respectively. The thresholds values are 9205~e and 9130~e for n- and p-side, respectively, and their threshold spread of 228~e and 180~e. 

During the series characterization of the E16-STS modules, different thresholds were set in the calibration, initially high, to ensure better performance due to its dependence with the noise level. Lower threshold were also set to explore the feasibility of calibrating and operating the modules at lower thresholds, and evaluate the potential to achieve enhanced performance. 
The gain and threshold spread, within approximately 1\% and 2\% for each sensor side, are considered sufficient to demonstrate a proper calibration of the channels. The small residual spread after calibration indicates that channel-to-channel variations are effectively compensated, resulting in a sufficiently uniform response across the channels.

\subsubsection{Measurement of the module’s noise levels}
The equivalent noise charge (ENC) of each module is extracted from threshold-scan measurements. Electronic noise causes the discriminator efficiency to vary smoothly around the effective threshold, resulting in the characteristic S-curve response. The width of this transition region is directly related to the noise level and is therefore used to determine the ENC.

\begin{figure}[h] 
    \centering
    \includegraphics[width=0.85\textwidth]{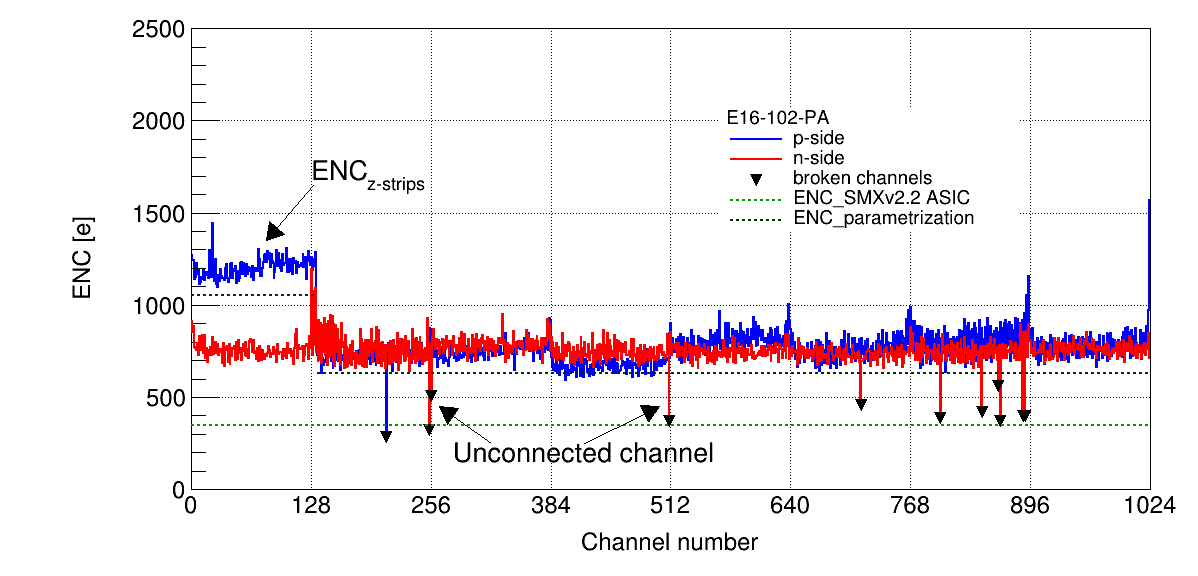}
    \caption{ENC measurement across all channels for the E16-102-PA module. The red and blue lines represent the n- and p-side channels, respectively. The green dashed lines indicate the expected ENC from the SMX ASIC and the ENC parametrization. Unconnected channels are marked with black triangles. The higher ENC observed for the Z-strips is also labeled.}
    \label{fig:ENC_e16_102_pa}
\end{figure}

Figure~\ref{fig:ENC_e16_102_pa} shows the distribution of the measured ENC for an exemplary module (E16-102-PA). The mean ENC agrees within 20\% with the analytically calculated value based on the total module capacitance \cite{RodriguezRodriguez2024}, taking into account the contribution of all individual module components. This level of agreement validates the capacitance model and indicates a good understanding of the detector noise characteristics. 

Few differences between odd-even channels are related with the topology of the signal lines and the position adopted by the microcables. 
A larger value of ENC is visible in the first group of channels on the p-side. These strips are those at the edges of the sensor p-side, which are interconnected via a double-metalization layer, making their length and overall capacitance larger than regular strips, and in consequence the increasing of the noise for this group of channels. They are commonly referred as Z-strips.

Unconnected channels can be recognized by their reduced noise compared to the connected ones, and are categorized based on noise levels: No-Analog-Response (NAR), where ENC is 0 and the channel is permanently damaged; broken TAB bond at the ASIC, causing a 350 e noise level, due to a faulty ASIC-microcable connection; and broken TAB bond at the sensor, where noise varies depending on the module’s average ENC and neighboring strips. Table \ref{tab:table_summary} summarizes the results of the  characterization for all the modules of the E16-STS setup.

Four modules, damaged during the commissioning run in November 2023 and the beam campaign in May 2024 were refurbished, characterized, and also integrated on ladders. Table~\ref{tab:table_summary} provides a summary of the parameters evaluated.
	
\begin{table}[h]
	\caption{Test results of the 10 + 4 E16-STS modules designated for the 2023 comissioning run and 2025 beam campaign. Every parameter is given as the mean value of the corresponding distribution, with the exception of the total of broken channels. The errors denote the standard deviation of the given values.}
    \centering
	\begin{tabular}{c c c c c c }
		\\ \hline
		\multicolumn{1}{l}{Module ID} & side & ENC & ADC gain & Threshold & Number of \\ \hfill
		 & & [e] & [e/LSB] & \multicolumn{1}{c}{[e]} & broken channels\\ \hline 
E16-101-PB & n & 960 $\pm$ 40 &  2095 $\pm$ 12 & \multicolumn{1}{r}{14464 $\pm$ 186} & \multicolumn{1}{c}{26} \\ \hfill
        ~~ & p & 730 $\pm$ 30 &  2098 $\pm$ 11 & \multicolumn{1}{r}{15011 $\pm$ 242} & \multicolumn{1}{c}{4}  \\
E16-102-PA & n & 940 $\pm$ 40 &  2098 $\pm$ 12 & \multicolumn{1}{r}{8944 $\pm$ 172} & \multicolumn{1}{c}{10} \\ \hfill   
        ~~ & p & 780 $\pm$ 25 &  2097 $\pm$ 13 & \multicolumn{1}{r}{9311 $\pm$ 169} & \multicolumn{1}{c}{1}  \\
E16-103-PB & n & 920 $\pm$ 40 &  2094 $\pm$ 13 & \multicolumn{1}{r}{14548 $\pm$ 177} & \multicolumn{1}{c}{7}  \\ \hfill
        ~~ & p & 770 $\pm$ 25 &  2096 $\pm$ 11 & \multicolumn{1}{r}{14530 $\pm$ 165} & \multicolumn{1}{c}{25} \\
E16-104-PA & n & 910 $\pm$ 35 &  2103 $\pm$ 12 & \multicolumn{1}{r}{14533 $\pm$ 224} & \multicolumn{1}{c}{1}  \\ \hfill 
        ~~ & p & 750 $\pm$ 25 &  2095 $\pm$ 11 & \multicolumn{1}{r}{14383 $\pm$ 167} & \multicolumn{1}{c}{4}  \\
E16-106-PB & n & 880 $\pm$ 35 &  2098 $\pm$ 13 & \multicolumn{1}{r}{9205 $\pm$ 180} & \multicolumn{1}{c}{3}  \\ \hfill 
        ~~ & p & 710 $\pm$ 25 &  2101 $\pm$ 13 & \multicolumn{1}{r}{9130 $\pm$ 228} & \multicolumn{1}{c}{2}  \\
E16-107-PA & n & 860 $\pm$ 40 &  2088 $\pm$ 14 & \multicolumn{1}{r}{9254 $\pm$ 193} & \multicolumn{1}{c}{1}  \\ \hfill 
        ~~ & p & 770 $\pm$ 30 &  2099 $\pm$ 12 & \multicolumn{1}{r}{9314 $\pm$ 212} & \multicolumn{1}{c}{19} \\
E16-108-PB & n & 840 $\pm$ 40 &  2100 $\pm$ 11 & \multicolumn{1}{r}{14823 $\pm$ 168} & \multicolumn{1}{c}{0}  \\ \hfill 
        ~~ & p & 800 $\pm$ 30 &  2096 $\pm$ 13 & \multicolumn{1}{r}{14611 $\pm$ 238} & \multicolumn{1}{c}{0}  \\
E16-109-PA & n & 820 $\pm$ 35 &  2094 $\pm$ 17 & \multicolumn{1}{r}{9615 $\pm$ 200} & \multicolumn{1}{c}{7}  \\ \hfill
        ~~ & p & 790 $\pm$ 30 &  2098 $\pm$ 12 & \multicolumn{1}{r}{9253 $\pm$ 181} & \multicolumn{1}{c}{8}  \\
E16-206-PA & n & 760 $\pm$ 39 &  2093 $\pm$ 12 & \multicolumn{1}{r}{9339 $\pm$ 159} & \multicolumn{1}{c}{7}  \\ \hfill 
        ~~ & p & 770 $\pm$ 30 &  2095 $\pm$ 14 & \multicolumn{1}{r}{9266 $\pm$ 200} & \multicolumn{1}{c}{9}  \\
E16-207-PA & n & 700 $\pm$ 30 &  2092 $\pm$ 10 & \multicolumn{1}{r}{9306 $\pm$ 141} & \multicolumn{1}{c}{1}  \\ \hfill
        ~~ & p & 780 $\pm$ 25 &  2096 $\pm$ 11 & \multicolumn{1}{r}{9290 $\pm$ 175} & \multicolumn{1}{c}{17} \\
\\ \hline \\
E16-101-PBv2 & n & 770 $\pm$ 39 & 2104 $\pm$ 11 & \multicolumn{1}{r}{14192 $\pm$ 157}  & 3 \\ \hfill
          ~~ & p & 726 $\pm$ 43 & 2105 $\pm$ 14 & \multicolumn{1}{r}{14186 $\pm$ 147} & 8  \\
E16-103-PBv2 & n & 799 $\pm$ 42 & 2105 $\pm$ 15 & \multicolumn{1}{r}{14205 $\pm$ 180} & 1  \\ \hfill
          ~~ & p & 699 $\pm$ 39 & 2104 $\pm$ 15 & \multicolumn{1}{r}{14168 $\pm$ 157} & 0 \\
E16-104-PAv2 & n & 741 $\pm$ 45 & 2095 $\pm$ 12 & \multicolumn{1}{r}{14165 $\pm$ 162} & 1  \\ \hfill 
          ~~ & p & 706 $\pm$ 44 & 2104 $\pm$ 14 & \multicolumn{1}{r}{14161 $\pm$ 155} & 0  \\
E16-106-PBv2 & n & 735 $\pm$ 36 & 2092 $\pm$ 18 & \multicolumn{1}{r}{14191 $\pm$ 180} & 0  \\ \hfill 
          ~~ & p & 693 $\pm$ 39 & 2099 $\pm$ 14 & \multicolumn{1}{r}{14140 $\pm$ 228} & 0  \\
	\\ \hline
	\end{tabular}
	\label{tab:table_summary}
\end{table}

\subsection{Response to a radioactive source}
\begin{figure}[h] 
    \centering
    \includegraphics[width=0.85\textwidth]{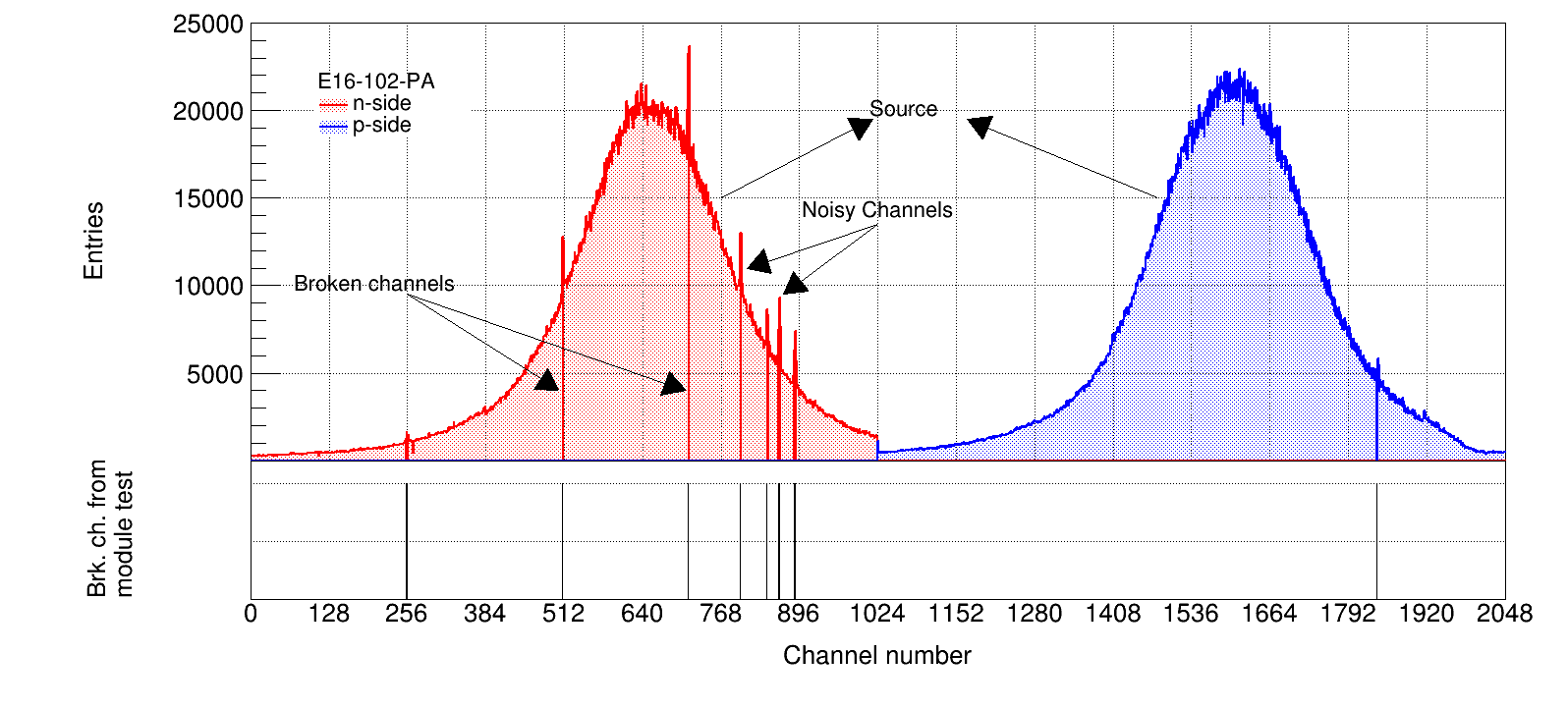}
    \caption{Channel distribution of the module E16-102-PA resulted of the irradiation of the sensor with ${}^{90}Sr/Y$. The broken channels identified through an ENC measurement are also displayed.}
    \label{fig:ch_dist_e16_102_pa}
\end{figure}

The signal readout capabilities of the modules are evaluated using a collimated $\beta$-emitting radioactive source, specifically ${}^{90}Sr/Y$, which emits electrons up to 2.2~\si{\mega\electronvolt}. 
The channel distribution of the signal is shown in Figure \ref{fig:ch_dist_e16_102_pa} for a source placed at 5~\si{\centi\meter} from the module. The first 1024 channels correspond to the strips in the n-side, while the last 1024 to the p-side. The two peaks around the central strips, correspond to the location where the source illuminated the sensor. Broken channels confirm the effectiveness of the identification in the ENC measurement. 

\begin{figure}[h!]
    \centering
    \begin{multicols}{2}
        \includegraphics[width=\linewidth]{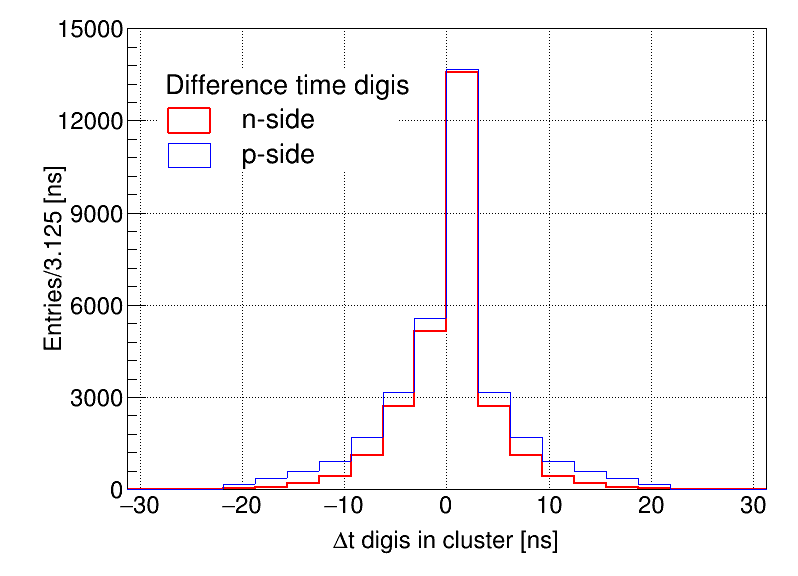} \par
        \includegraphics[width=\linewidth]{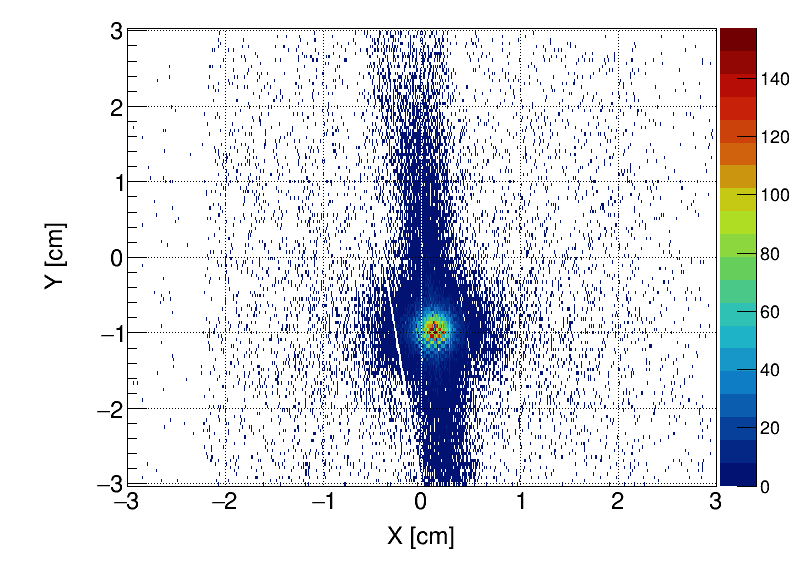} \par
    \end{multicols}
    \caption{Left:  Time difference for strip-signals inside the cluster, demonstrating the channel synchronization; Right: XY hit reconstruction of a radioactive ${}^{90}Sr/Y$ source illuminating in the center of the module.}
    \label{fig:e1_time_xy}
\end{figure}

The signals from neighboring strips are grouped together to form a cluster. Figure~\ref{fig:e1_time_xy} left panel shows the distribution of the time differences between the signal strips within a cluster; the clear peak around zero demonstrates their synchronization and justifies the 25~\si{\nano\second} time window for the reconstruction of clusters, and subsequently of hits.
Hits are then formed by correlating a cluster on the p-side and one on the n-side within the same time window.

The geometrical space distribution of the reconstructed hits (space-points) is shown in Figure \ref{fig:e1_time_xy} right.
The reconstructed 2D hit positions show a concentration at the expected source location, consistent with a collimated $\beta$ source positioned close to the sensor,
validating both the module's response and the reconstruction software. 

\section{Commissioning and Operation in beam test experiments}

\begin{figure}[h] 
    \centering
    \includegraphics[width=0.9\textwidth]{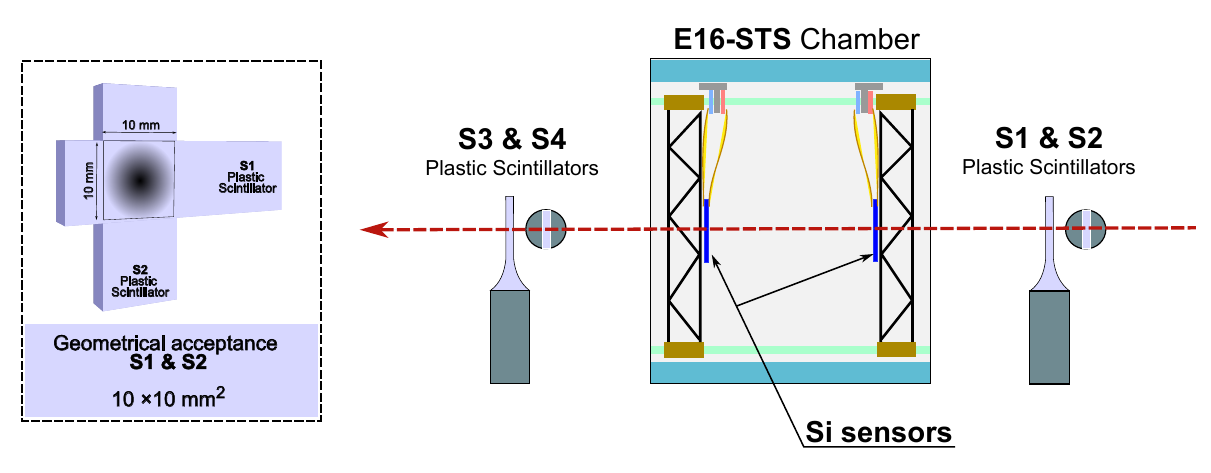}
    \caption{Experimental setup during November 2023 beam campaign at KEK, using an electron beam of 3~\si{\giga\electronvolt}/c momentum. The setup includes the E16-STS chamber with a combination of alignment and trigger scintillators used to define an active area of 10~$\times$~10~\si{\milli\meter}$^2$.}
    \label{fig:beamline}
\end{figure}
In November 2023 a beam test experiment was conducted at the KEK PF-AR  beamline intermittently exposing the STS modules to an electron beam with a momentum of 3~\si{\giga\electronvolt}/c. The  E16-STS chamber was sandwiched between four scintillators, each measuring 10~$\times$~10~\si{\milli\meter}$^2$, as shown in Figure~\ref{fig:beamline}. Two pairs of scintillators were positioned in front and behind of the chamber.
The setup was designed to evaluate a small spot on the sensor. The sensors were
positioned approximately 3~\si{\meter} from the beam extraction point, with a distance of 1.5~\si{\meter} between
the two sets of scintillators. The scintillator irradiation position and the
central position of the beam were nearly aligned with the center of the sensor.
The coincidence signal from the four scintillators
was processed using a NIM coincidence module and routed to a standalone FEB
to record the scintillator timing. The average coincidence rate was 20~\si{\hertz}. 
Three out of the 10 modules installed in the E16-STS chamber were tested.
The noise level was 800~e (0.13~\si{\femto\coulomb}). The corresponding thresholds were set to 1.4~\si{\femto\coulomb} for the amplitude measuring path and 3.0~\si{\femto\coulomb} for the timing measuring path, although the latter was later found to be too higher, leading to a fraction of hits with wrong timing information.

\begin{figure}[h] 
    \centering
    \includegraphics[width=0.9\textwidth]{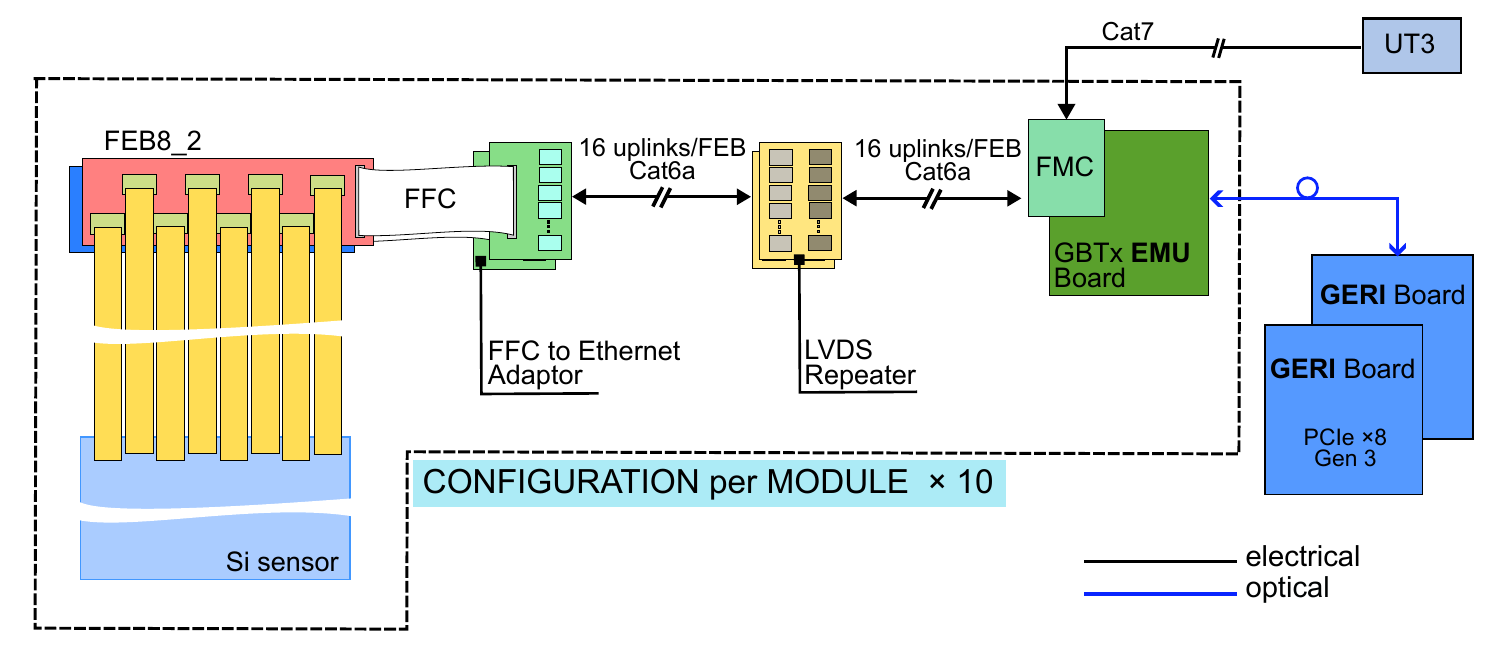}
    \caption{Schematic readout of the J-PARC E16-STS. The raw signal is sent from the sensor to GERI board through the FCC-Ethernet adapter, LVDS repeater and GBTx-EMU board.}
    \label{fig:readout}
\end{figure}

A functional block diagram of the readout chain of the STS system used for the test beam is schematically depicted in Figure~\ref{fig:readout}. The analog charge deposited on the strips is amplified, shaped and digitized in the SMX ASICs on the FEB. Digital hit frames are transmitted via  flexible flat cables, interfaced via six LVDS repeaters (signal amplifying relays) to Cat6a cables for long-distance transmission. During the test-beam experiment, however, the LVDS repeaters were removed from the setup, since
they introduced communication instabilities and additional noise under the specific operating conditions of the test.
The signal is finally transmitted to the GBTx Emulator (GBTxEMU) board, which receives the data from multiple uplinks via a custom Mezzanine Card (FMC) and aggregates the data for optical transmission. Data is further transmitted via optical fibers to a PCI-Gen3x8 board (GERI), originally developed for the Silicon Tracker of the BM@N experiment \cite{GERI} and customized for the
E16 experiment. The GERI board integrates information from all sensors (up to seven sensors per GERI board) and
transmits it to the computer’s memory using Direct Memory Access (DMA) mode. It utilizes a GBT-FPGA based firmware
for data transfer, slow control, timing and trigger control. 
The system is powered by multiple LV modules and one HV
module, while temperature control is ensured through a cooling plate connected to a Yamato CF303 chiller \cite{yamato}. During
this test the system operated in free streaming mode. The timestamp in all ASICs of the system were synchronized
before data taking.

\begin{figure}[h!]
    \centering
    \begin{multicols}{3}
        \includegraphics[width=\linewidth]{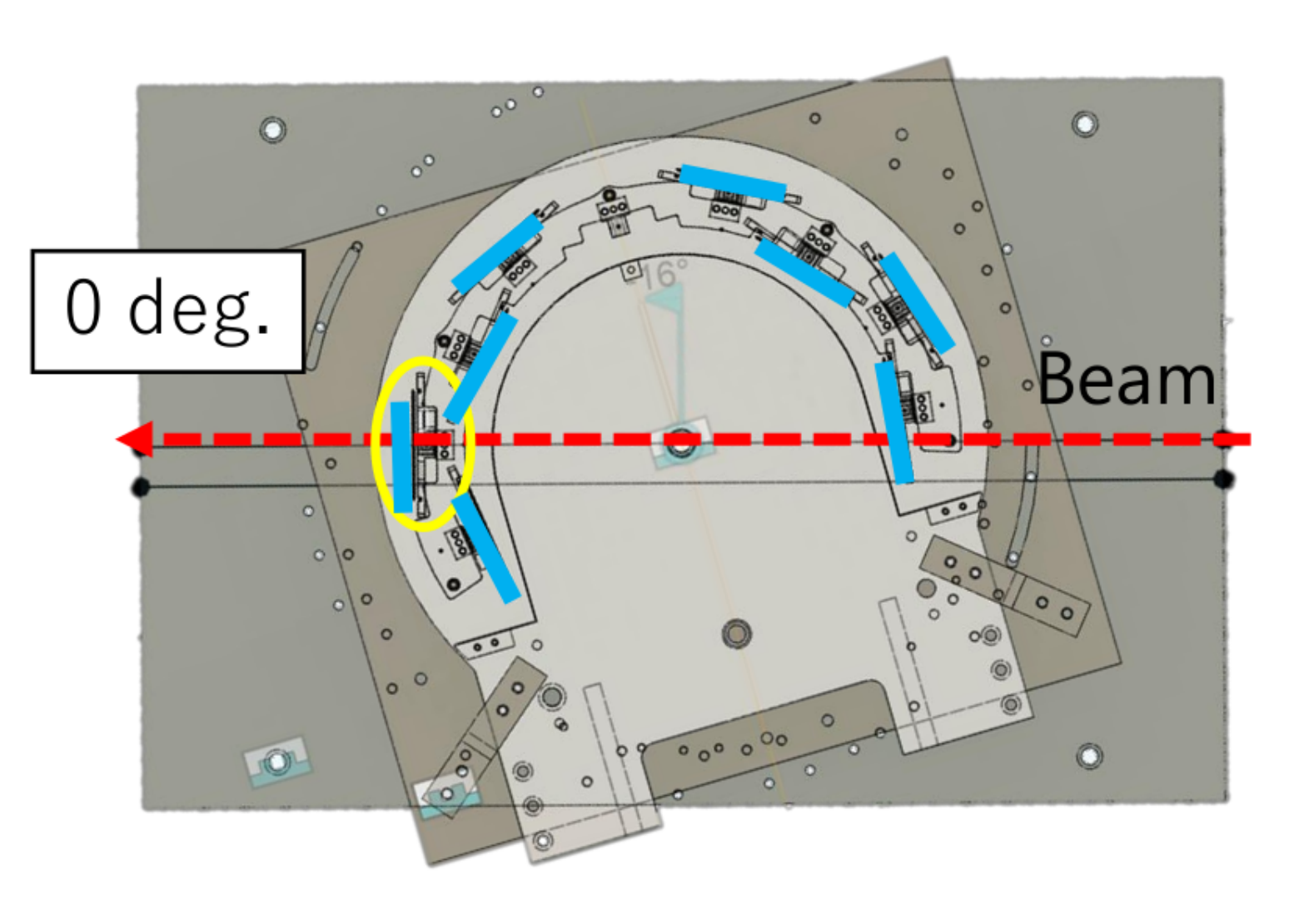} \par
        \subcaption[]{${0}^{~\circ}$}
        \includegraphics[width=\linewidth]{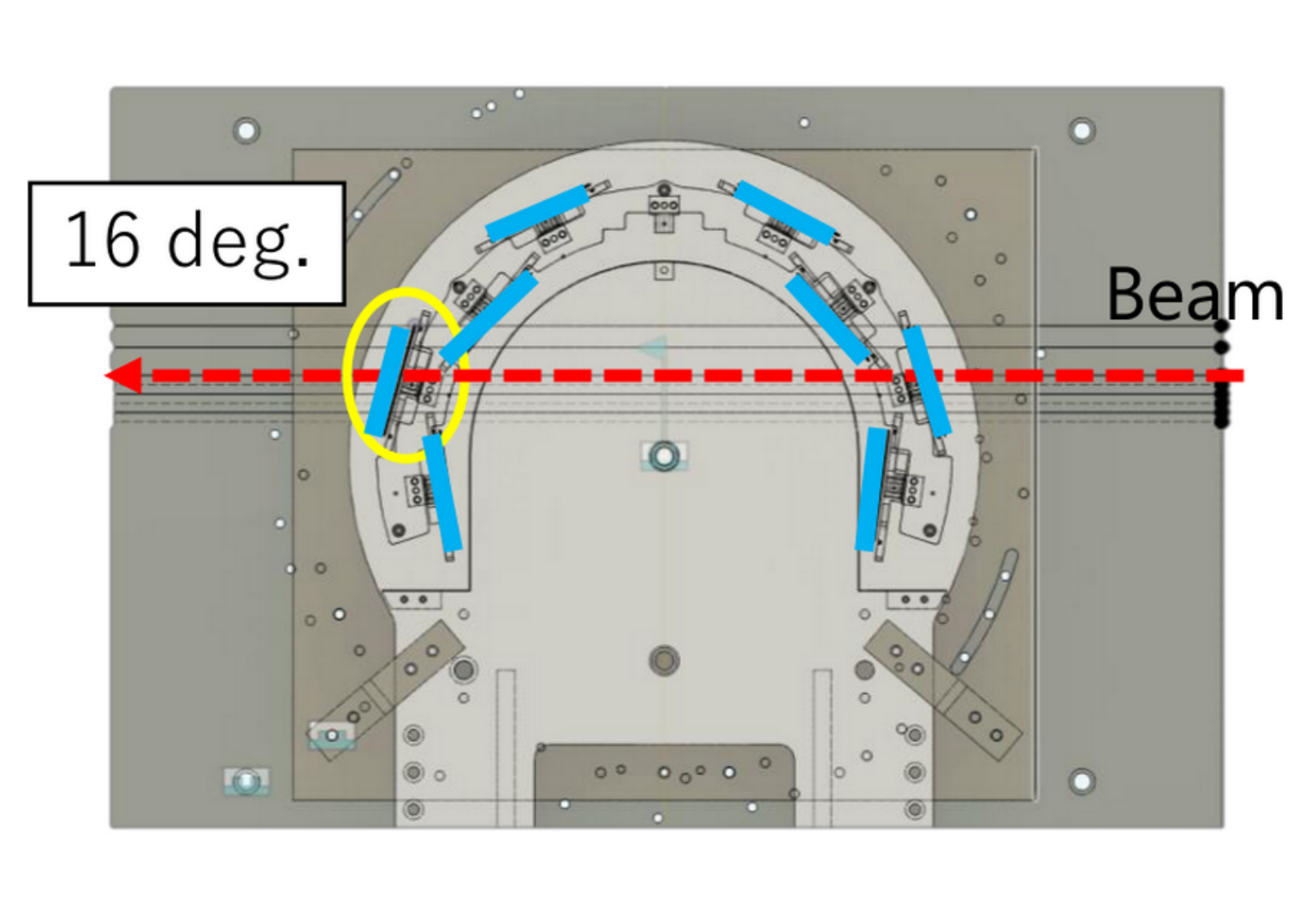} \par
        \subcaption[]{${16}^{~\circ}$}
        \includegraphics[width=0.93\linewidth]{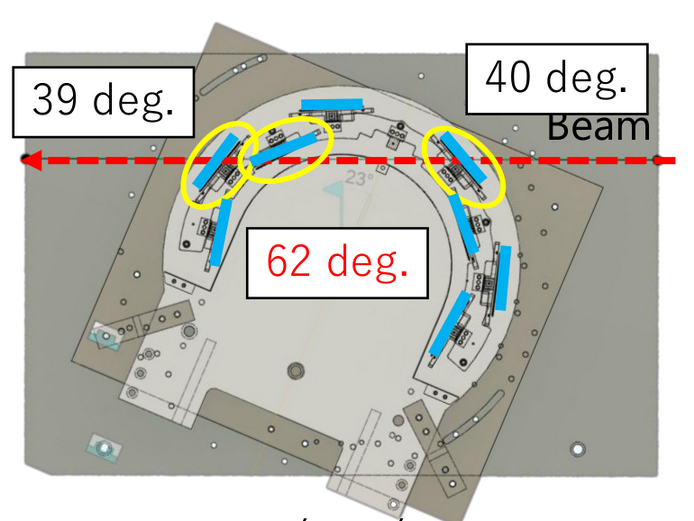} \par
        \subcaption[]{different angles}
    \end{multicols}
    \caption{Two different testing configurations. Module E16-108-PA positioned at an incident angle of: a) ${0}^{\circ}$ and b) ${16}^{\circ}$. c) The three modules at different angles of incidence.}
    \label{fig:config_setup}
\end{figure}

The modules were tested in two configurations, shown in Figure~\ref{fig:config_setup}. In the first, data was recorded from a single module while varying the beam incidence angles (0$^\circ$ and 16$^\circ$) to assess the response to different beam orientations. In the second configuration, three modules recorded data simultaneously, each with different incidence angles. 

\section{Results}
\subsection{Data Analysis}
The strategy followed for data analysis is to use the reconstruction chain developed for the CBM-STS detector, therefore interfacing the data at the earliest possible stage (raw data) to the CBMROOT format \cite{AlTurany2006CbmRoot}, and thereafter treat them with CBM software.

The raw data are compacted into a series of 32-bit words by the GERI board. The data stream contains timing information from both the GERI board and the ASIC hit frame, including channel-address mapping from the e-link, a timestamp value of the signal expanded beyond the SMX internal timestamp, 
and the associated charge information. The hit time is provided by the SMX ASIC via two independent data frames as the Least Significant Bits (LSB) and the Most Significant Bits (MSB) of the timestamp.
Two overlapping bits ensure correct matching of LSB and MSB, and consistent linear progression of time. The time
information is further extended by the backend FPGA on the GERI board to provide the absolute time. The data are
stored in a CBMROOT-tree which contains the information of the raw signals, i.e. address (which identifies the module),
channel, amplitude and time. It is organized per Time Slices (TS), i.e. discrete time intervals of data. The duration of
the TS is set by a fixed number of the periodic reset signals of the GERI board.

Clusters are reconstructed by correlating signals from adjacent activated strips within a small time window of 25~\si{\nano\second} \cite{Friese2019}. The amplitudes of the signals from each fired strip are summed, and the time information is averaged. The cluster's center of gravity is then used as an estimate for the particle's crossing point in the detector's sensitive volume \cite{Malygina_2017}. Finally, hits, or space points, are obtained by correlating the p-side and n-side clusters, and geometrical coordinates are assigned using a CBMROOT-based description of the setup.

\subsection{Signal amplitude}

\begin{figure}[h!]
    \begin{subfigure}{0.32\textwidth}
       \includegraphics[width=\textwidth]{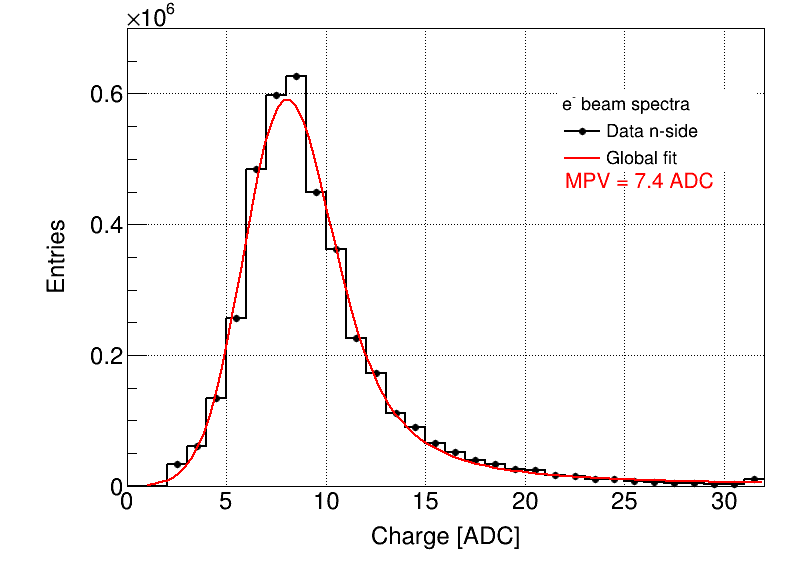}
       \caption{${0}^{\circ}$ n-side}
    \end{subfigure}\hfill
    \begin{subfigure}{0.32\textwidth}
       \includegraphics[width=\textwidth]{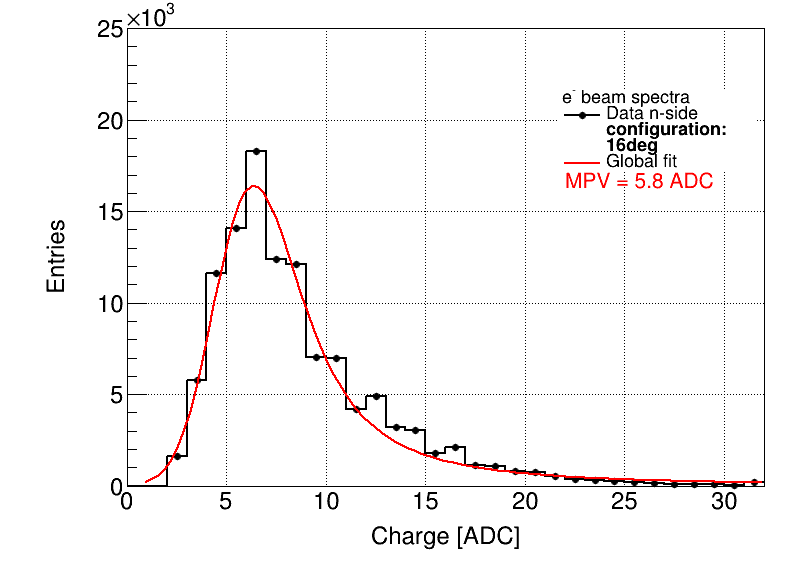}
       \caption{${16}^{\circ}$ n-side}
    \end{subfigure}\hfill
    \begin{subfigure}{0.32\textwidth}
       \includegraphics[width=\textwidth]{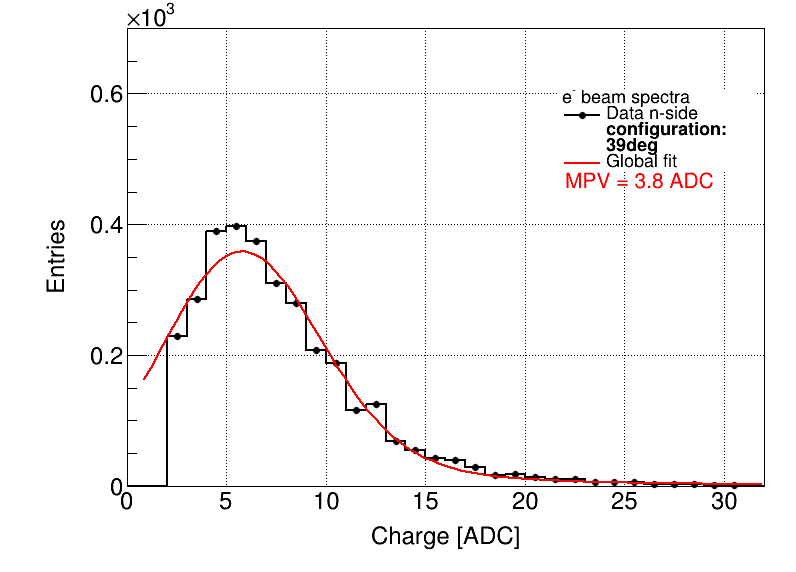}
       \caption{${39}^{\circ}$ n-side}
    \end{subfigure}

    \bigskip 
    \begin{subfigure}{0.32\textwidth}
       \includegraphics[width=\textwidth]{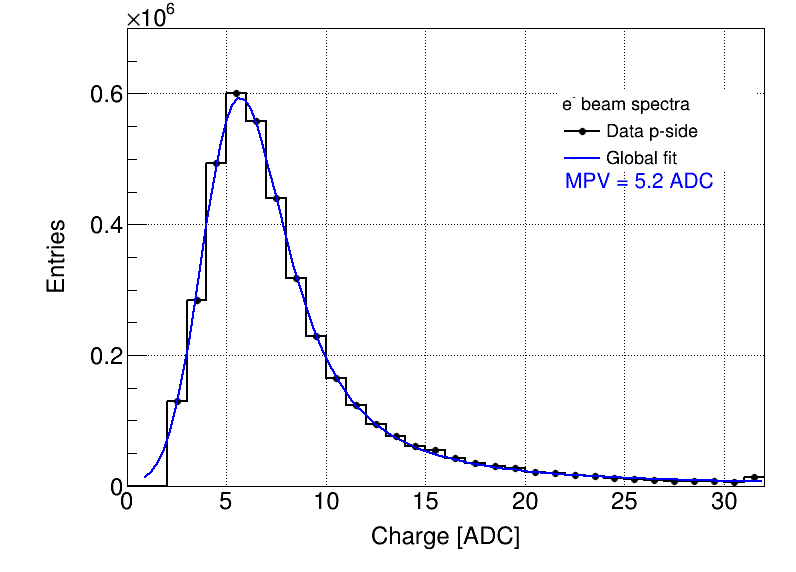}
       \caption{${0}^{\circ}$ p-side}
    \end{subfigure}\hfill
    \begin{subfigure}{0.32\textwidth}
       \includegraphics[width=\textwidth]{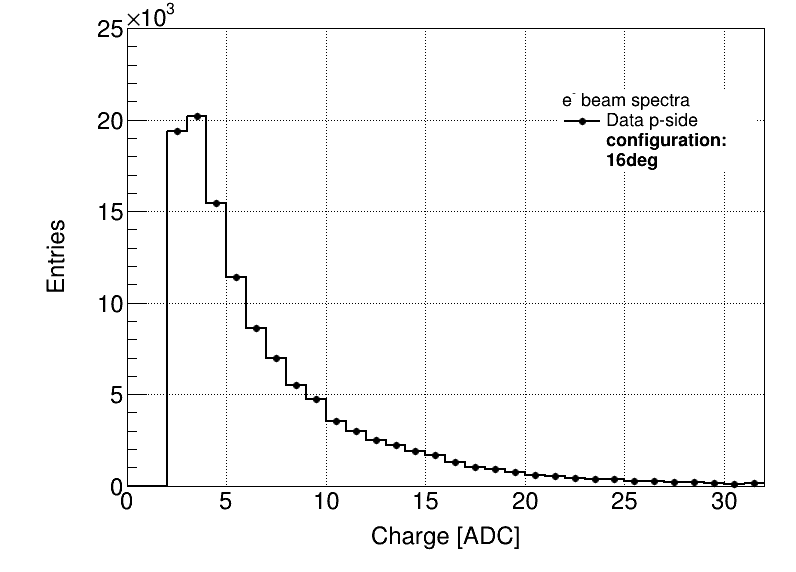}
       \caption{${16}^{\circ}$ p-side}
    \end{subfigure}\hfill
    \begin{subfigure}{0.32\textwidth}
       \includegraphics[width=\textwidth]{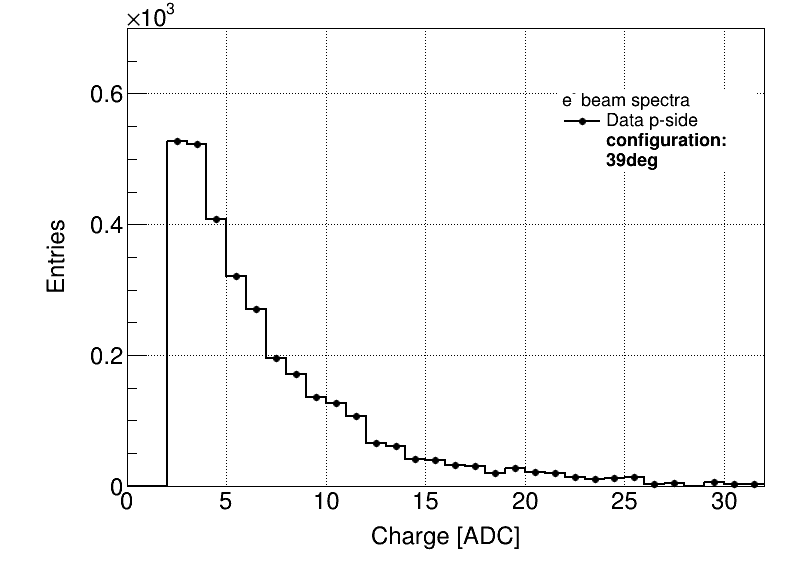}
       \caption{${39}^{\circ}$ p-side}
    \end{subfigure}
        \caption{Distribution of signal amplitude for the n- and p- side of a module with inclination angle of 0$^\circ$, 16$^\circ$ and 39$^\circ$. The data distributions are fitted with a convolution of a Landau function with a Gaussian.}
    \label{fig:charge}
    \end{figure}

To evaluate the signal-to-noise ratio (S/N), the measured signal amplitude is compared with noise measurements obtained during module characterization. The signal amplitude (S) is determined by the most probable value (MPV) of the charge distribution, shown in Figure~\ref{fig:charge} for the same module at different inclination angles. 

As the modules operated with relatively high thresholds (approximately 8750~e and 11000~e, for the n- and p-side respectively), only small contribution of noise is visible at low amplitudes.
The average signal (S) is obtained by fitting the distribution with a Landau function convoluted with a Gaussian, yielding a most probable value (MPV) of 22212~e~$\pm$~980~e for the n-
side and 19803~e~$\pm$~520~e for p-side. The uncertainty was estimated accounting for the ADC gain and threshold spread, as well as the systematic error of the fit, with the statistical error being negligible. 
The measured values are up to 7\% below the expected charge deposited by minimum ionizing particles (MIPs) in 320~\si{\micro\meter} of silicon, which can be explained with signal losses due to high thresholds and the distribution of charge over multiple strips, with low-amplitude strips being cut out.
Using the corresponding ENC noise measurements of 795~e and 695~e, indicated in Table \ref{tab:table_summary}, an average signal-to-noise ratio (S/N) of $27.9~\pm~1.5$ and $28.5~\pm~1.5$ was determined for the n- and p-side, respectively.
    
The middle and right panels of the Figure~\ref{fig:charge} show the charge distribution for inclined tracks across the detector. As the incident angle of the tracks increases, a more pronounced charge sharing effect between adjacent strips is observed, confirmed by the average cluster size which increases from 1.1 to 1.6 as the incidence angle increases from 0$^\circ$ to 39$^\circ$. This behavior is attributed to the longer path length of the particle through the 320~\si{\micro\meter} thick sensor material, combined with the narrow strip pitch of only 58~\si{\micro\meter}, which causes the deposited charge to be distributed over a larger number of strips. Consequently, the charge collected by a single strip decreases, leading to an observable shift of the distribution peaks toward lower ADC values.

\begin{figure}[h!]
    \centering
    \begin{multicols}{3}
        \includegraphics[width=\linewidth]{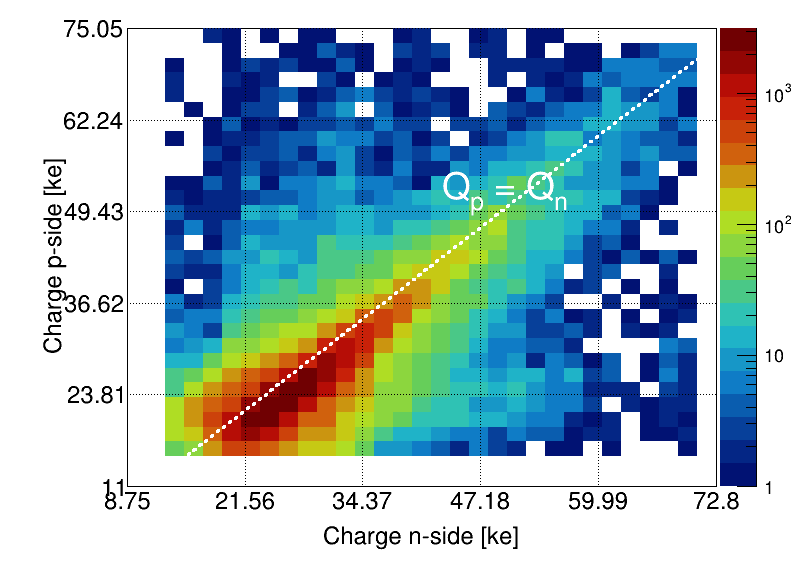} \par
        \subcaption[]{${0}^{\circ}$}
        \includegraphics[width=\linewidth]{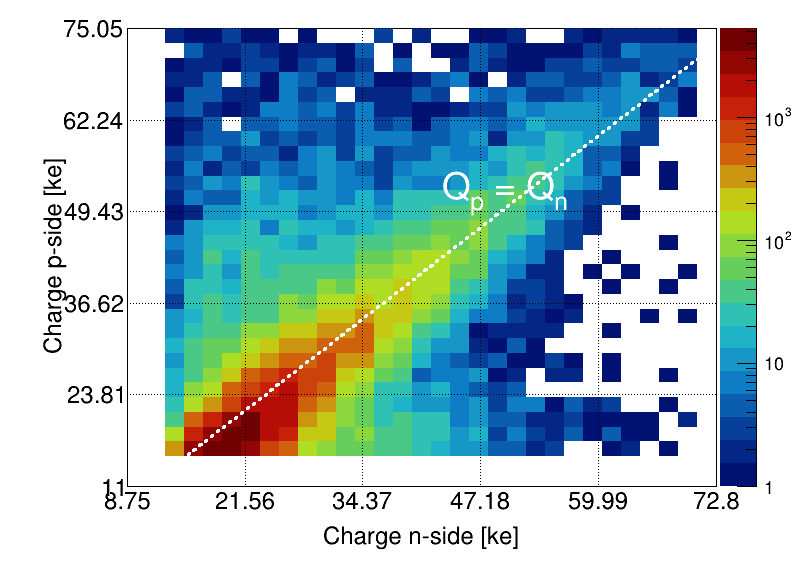} \par
        \subcaption[]{${16}^{\circ}$}
        \includegraphics[width=\linewidth]{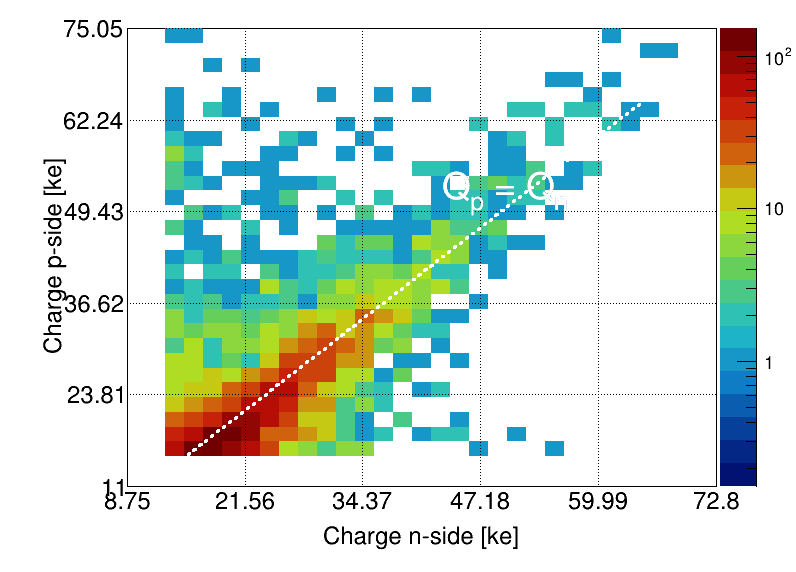} \par
        \subcaption[]{${39}^{\circ}$}
    \end{multicols}
    \caption{Signal amplitude correlation between n- and p-side for clusters of size 1. Data measured  for the different testing configurations of incidence angle.}
    \label{fig:runs_pn_correlation}
\end{figure}

Figure~\ref{fig:runs_pn_correlation} shows the correlation of signal amplitudes of n- and p-side. A strong correlation is observed, around 23~ke, which corresponds to the MPV of the charge distribution and is very close to the deposited charge by MIPs. The difference in the axis scales between the n- and p-side is due to differences in threshold settings applied during the data taking. 
In the other configurations shown in the Figure~\ref{fig:runs_pn_correlation}, the correlation remains clearly visible, though the MPV shifts to lower charge values, due to the increased sharing of charge. 

\subsection{Time Resolution}
Accurate timing calibration is crucial for establishing reliable time correlations between signals linked to the same physical event. It is therefore necessary to adjust the offsets of the ASICs individually, using a common reference detector with excellent time resolution. The reference here is taken from the scintillators. 

\begin{figure}[h!]
   \centering
        \includegraphics[width=\textwidth]{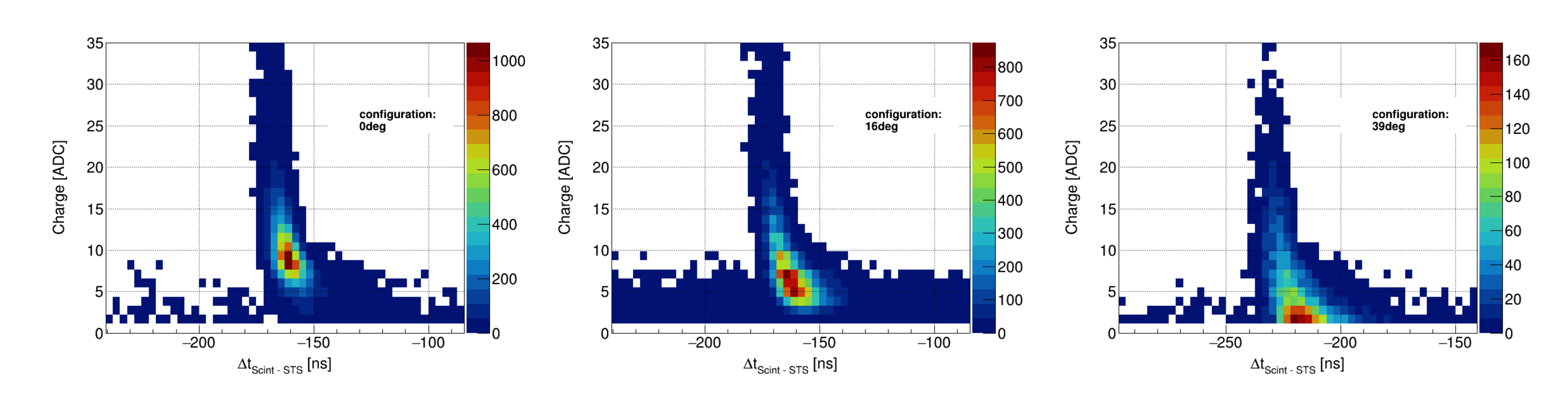} 
        \includegraphics[width=\textwidth]{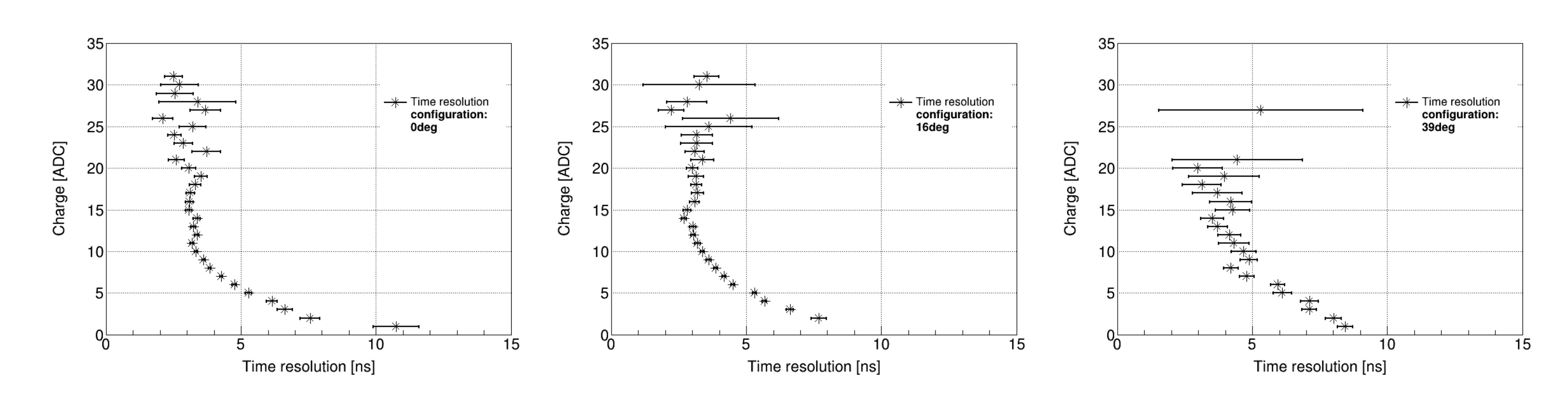} 
        \caption{Top: Time difference of raw strip signals of the STS and signal from the scintillators as a function of the STS signal amplitude for $0^{\circ}$ (left), $16^{\circ}$ (middle) and $39^{\circ}$ (right) inclination angle. Bottom: Time resolution of the STS, as a function of the STS signal amplitude for the same inclination angles.}
    \label{fig:time}
\end{figure}

Figure~\ref{fig:time} -top panels- shows the time difference between the signals from the STS module and the scintillators as a function of the STS signal's amplitude. The average time difference of around 160~\si{\nano\second} reflects a time offset between the systems, caused by differing propagation times in the readout chain. 

Additionally, the Figure~\ref{fig:time} shows a delay for signals with lower amplitudes (time walk effect) due to a fixed threshold and finite signal rise time.
The time calibration accounts for these effects. The resolution of the STS detector can be extracted from the width of the distribution, since the contribution from the scintillators is negligible (picoseconds). As shown in the bottom panels of the Figure~\ref{fig:time}, for signal amplitude above 10 ADC (corresponding to $\sim~$29000~e) the standard deviation is 3.2~\si{\nano\second}, as expected from the ASIC specifications.

For the increased inclination angle, shown in the middle and right panels of Figure~\ref{fig:time}, the main signal peak shifts towards lower amplitudes due to the increased signal sharing, while the time resolution remains unchanged, reaching values of about 3.2~\si{\nano\second} for the highest signal amplitudes.

\subsection{Position resolution}

Using the data collected in the three-module configuration, the position resolution was studied, exploiting the fact that the beam traverses all three sensor planes sequentially. 
Tracks are first reconstructed by selecting the best combination of hits across the three sensors, using both spatial and timing information.  

\begin{figure}[h] 
    \centering
    \includegraphics[width=0.45\textwidth]{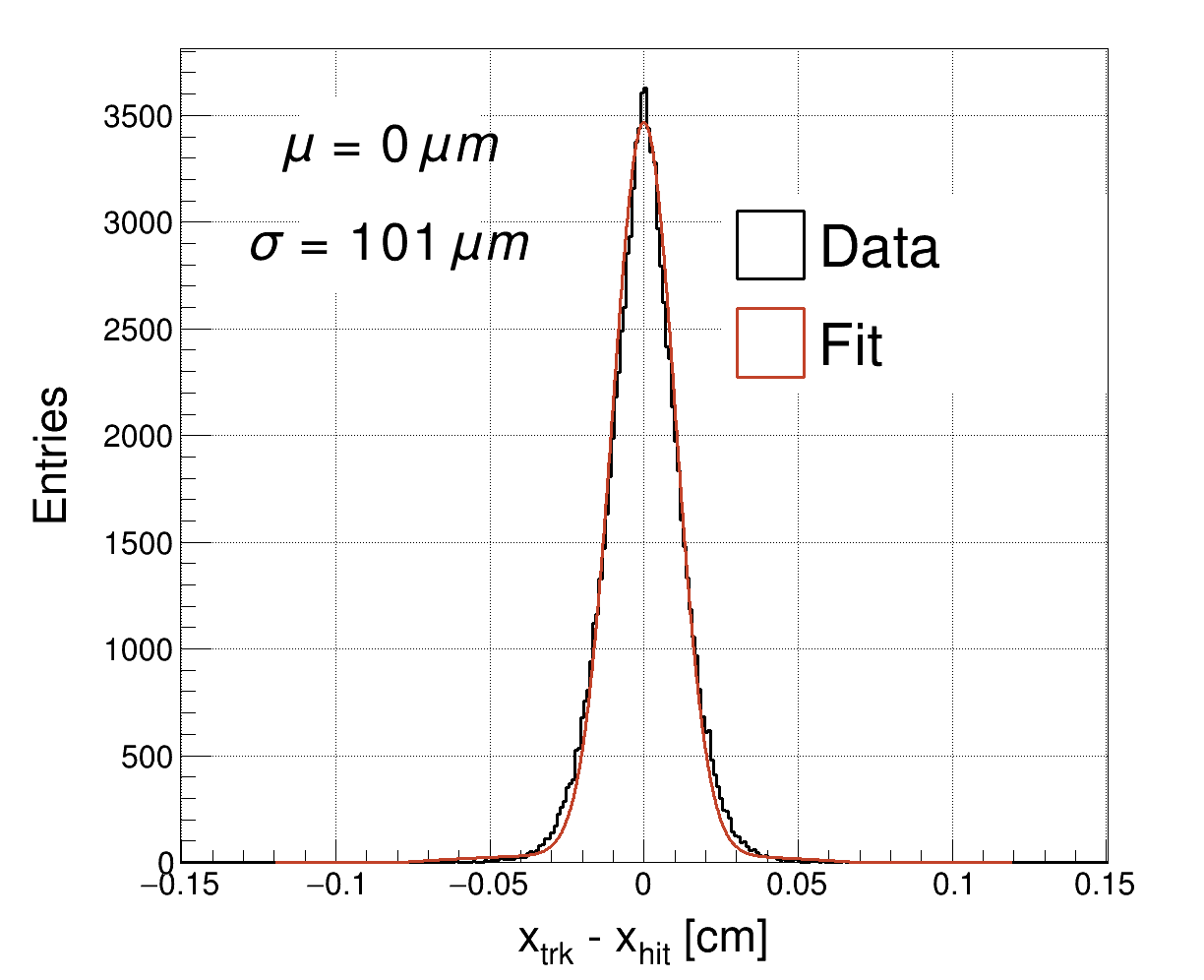}
    \includegraphics[width=0.45\textwidth]{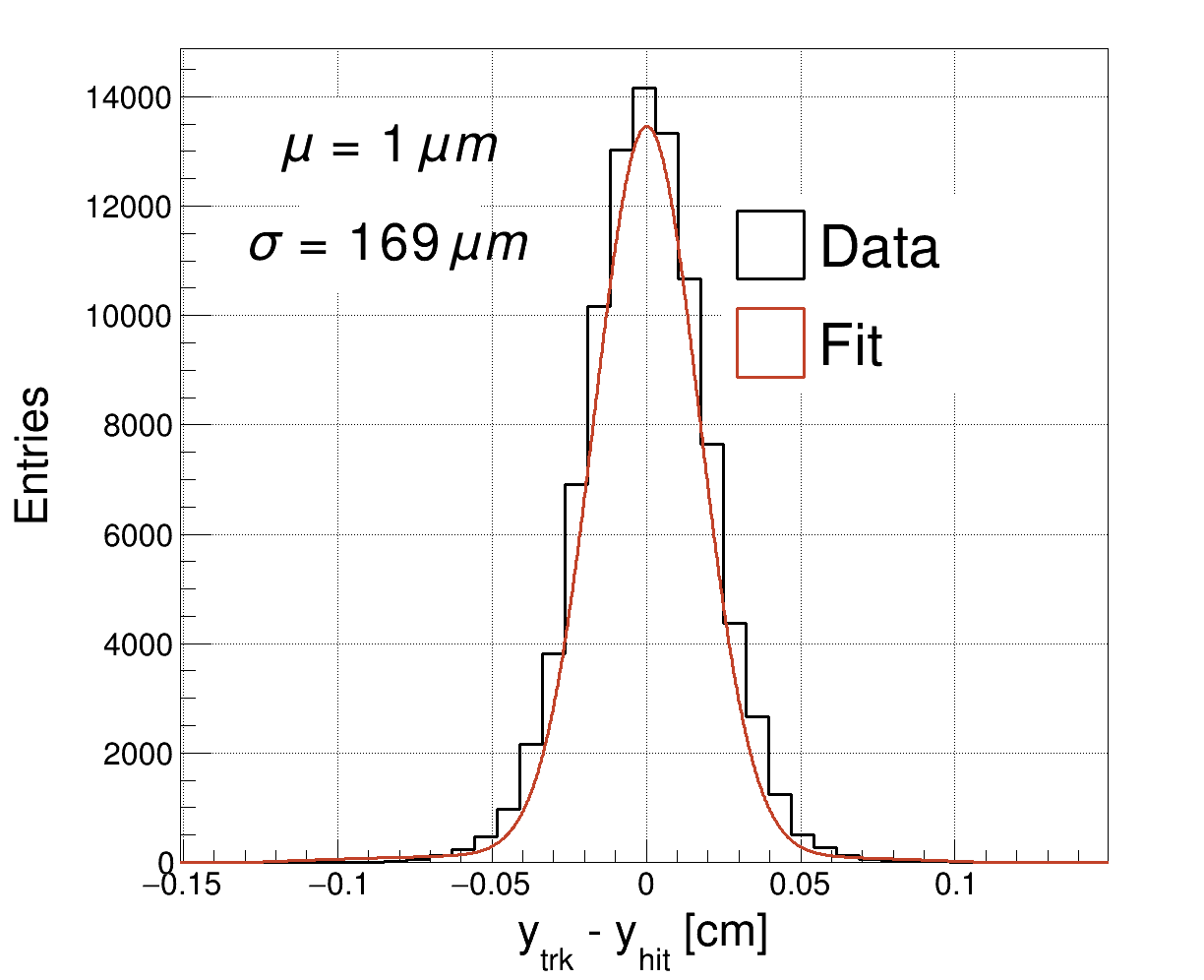}
    \caption{Residual distributions in the $X$ (left) and $Y$ (right) directions for tracks extrapolated to the middle sensor (DUT) at an inclination angle of $62^\circ$. The distribution is fitted with the sum of a Gaussian and a polynomial, to account for the background.}
    \label{fig:residual_xy}
\end{figure}

For each track candidate, a track segment is then constructed using hits from two sensor planes only, while the third plane is treated as the Device Under Test (DUT). 
Assuming a straight-line trajectory, the expected hit position on the DUT is obtained by extrapolating the track segment. 
Unbiased residuals in the $X$ and $Y$ directions are defined as the difference between the measured hit position on the DUT and the extrapolated track position. 
Figure~\ref{fig:residual_xy} shows the residual distributions in $X$ and $Y$ when the middle sensor is used as the DUT.

The intrinsic spatial resolution of the STS is extracted from the width of the residual distributions. 
The residuals are fitted with the sum of a Gaussian function and a polynomial, to account for the background, which provide a reasonable description of the observed shapes. 
The fitted means are close to zero, indicating negligible systematic offsets. 
The widths of the distributions, as shown in Fig.~\ref{fig:residual_xy}, are $113~\mu\mathrm{m}$ in $X$ and $192~\mu\mathrm{m}$ in $Y$.
These widths reflect the combined measurement uncertainty, arising from several contributions: the intrinsic detector resolution of the DUT, the track extrapolation uncertainty, and multiple scattering (the latter being neglected in the present treatment) and 
can therefore be expressed as 

\begin{equation}
\sigma_{\mathrm{res}}^2 = \sigma_{\mathrm{track}}^2 + \sigma_u^2
\label{eq:sigma_res}
\end{equation}
where $\sigma_u$ denotes the intrinsic detector resolution and $\sigma_{\mathrm{track}}$ the track prediction uncertainty at the DUT position.

The track extrapolation uncertainty depends on the resolutions of the tracking planes as well as on the detector geometry. 
In particular, due to the inclination of the sensors, the projection of the intrinsic resolution into the laboratory frame introduces a non-trivial dependence on the track angles and hit positions. 
More specifically, the inclination of the sensors in the $xz$ plane modifies the detector resolution projected onto the laboratory coordinates according to

\[
\sigma_{u_x} = \frac{\sigma_x}{\cos\alpha}
\hspace{2cm}
\sigma_{u_z} = \frac{\sigma_z}{\sin\alpha}
\]

where $\alpha$ is the angle between the particle trajectory (beam axis) and the sensor normal. 
As a consequence of the sensor inclination, the hit reconstruction also acquires a dependence on the $z$ coordinate, which therefore cannot be treated as constant and introduces an additional source of uncertainty in the track extrapolation.

For a straight-line track reconstructed from two measurements $i$ and $j$, the track uncertainty at the DUT can be written as
\begin{align}
\sigma_{\mathrm{track}}^2(x_{DUT}) &= \sigma_{x_i}^2
\left(\frac{z_{DUT} - z_j}{z_i - z_j}\right)^2 +
\sigma_{x_j}^2
\left(\frac{z_{DUT} - z_i}{z_i - z_j}\right)^2 \notag\\
&+
\sigma_{z_i}^2
\left(\frac{(x_i - x_j)(z_{DUT} - z_j)}{z_i - z_j}\right)^2
+
\sigma_{z_j}^2
\left(\frac{(x_i - x_j)(z_{DUT} - z_i)}{z_i - z_j}\right)^2
\label{eq:sigma_trk}
\end{align}
where $\sigma_{x}$ and $\sigma_{z}$ denote the projected measurement uncertainties in the laboratory frame. 

This expression highlights the non-trivial dependence of the track uncertainty on both the detector geometry and the inclination angles, implying that the extraction of the intrinsic resolution requires a proper evaluation of the track contribution on a track-by-track basis.
Substituting Eq.~\ref{eq:sigma_trk} into Eq.~\ref{eq:sigma_res}, the spatial resolution of the STS sensor, $\sigma_u$, is determined to be $(42~\pm~2)~\si{\micro\meter}$ and $(70~\pm~4)~\si{\micro\meter}$ in the $X$ and $Y$ directions, respectively, where the quoted systematic uncertainties reflect the maximum RMS values obtained for the resolution of all sensors.

Considering the large inclination angle, which enhances charge sharing, as well as the relatively high thresholds that may lead to partial signal loss and potential mis-reconstruction of hit positions, the achieved resolution is nevertheless consistent with expectations.

\section{Coupling the STS to a triggered system}
The SMX ASIC, developed for the free-streaming mode of the CBM experiment, operates in self-trigger mode, continuously sending all hit information to the Data Acquisition (DAQ) system. The STS system was run in free-streaming mode during the test beam-line commissioning. However, when coupling it to the externally triggered E16 setup for actual data-taking, the readout needed to be adapted to integrate with the triggered system.

To this end, an online trigger selection was implemented at the GERI stage, which transmits only the STS hits corresponding to the E16 triggers to the data acquisition for recording, greatly reducing the data volume and computing requirements.

\begin{figure}[h] 
    \centering
    \includegraphics[width=0.9\textwidth]{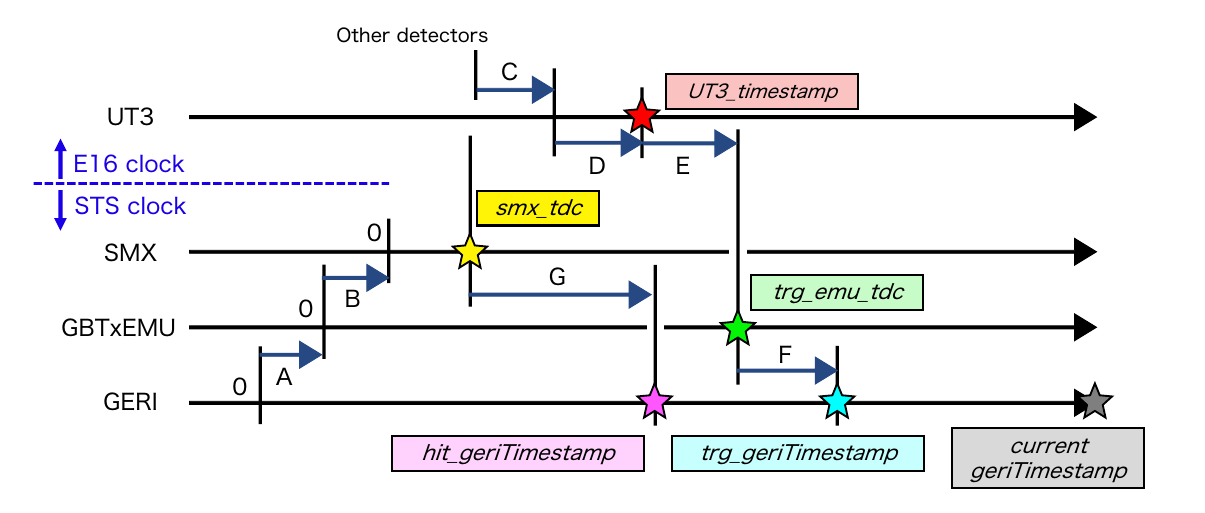}
    \caption{Schematic illustration of time axes and event timing in the STS readout system.}
    \label{fig:trigger}
\end{figure}

Implementing this online selection requires precise time synchronization among the STS readout components. Synchronization is typically classified into three levels: frequency synchronization, ensuring that the clock frequencies match; phase synchronization, aligning clock edges; and time synchronization, which aligns timestamps across all components and depends on accurate frequency and phase synchronization. In the STS readout chain, the GERI boards act as clock masters and supply a 40~\si{\mega\hertz} reference to all downstream components, which operate independently of the E16 DAQ clock. The GBTxEMU generates a 160~\si{\mega\hertz} clock based on the GERI reference and records a 64-bit timestamp, while the SMX utilizes the rising and falling edges of this 160~\si{\mega\hertz} clock to generate a 320~\si{\mega\hertz}, 14-bit timestamp.

Data selection near the trigger is performed by comparing timestamps from each component -SMX, GBTxEMU, GERI, and UT3- with the E16 trigger timestamp. Differences in these timestamps arise from signal path lengths, processing times, and trigger propagation delays. By correlating the timestamps, only hits coincident with the E16 trigger are selected for recording, as illustrated in Fig. \ref{fig:trigger}, enabling the integration of a free-streaming STS system with a triggered DAQ setup while keeping the data volume manageable.

\section{Summary}
The E16 experiment has been upgraded, replacing the original SSD modules with a lightweight, high-rate system based on the STS developed for the CBM experiment.
The E16-STS was built with pre-series modules of the CBM-STS that were assembled and characterized at GSI, transported to J-PARC and installed in the E16 experimental setup. 
Fifteen STS modules, including one prototype, ten pre-series, and four replacement units, were evaluated. The test covered IV curve measurements, ASIC calibration, electronic characterization, and source-based signal response. Results showed stable calibration, noise levels consistent with expectations, with all modules meeting production goals required for CBM \cite{ststdr2013} and E16 \cite{Yokkaichi2022}, and in-line with other productions of STS modules carried out recently \cite{Ramirez25}. A standardized testing protocol was established and is now applied to the final production of CBM-STS modules to ensure consistent quality and performance in series production.

The E16-STS chamber was operated in a beam test experiment at the KEK PF-AR  beamline with an electron beam of 3~\si{\giga\electronvolt}/c for commissioning the detector and assessing its performance. 
Although operating the detector at relatively large signal thresholds, the data collected allow to assess an excellent time resolution of 3.2~\si{\nano\second}, in line with the expectations from the ASIC characteristics.
The spatial resolutions, measured by exposing three sensors to the beam simultaneously, show the expected distinct performance and, despite large inclination angles, is consistent with design requirements. 
At all inclination angles, a signal-to-noise ratio above 25 was measured.

This confirms the ability of the STS modules to provide space point determination with the required resolution when used as the innermost detector layer, exposed to the highest particle flux with high intensity collision rate, thereby playing a crucial role in enhancing the experimental capability for precise tracking of the E16 experiment.

Furthermore, the adaptation of the free-streaming STS to the externally triggered E16 setup has been successfully implemented via online trigger selection and timestamp correlation, demonstrating the feasibility of integrating a self-triggering detector into a triggered DAQ environment without compromising performance.

\section*{Acknowledgements}
This work was carried out in the framework of the Arrangement for Research Cooperation in the Field of High-Intensity Proton Accelerators between GSI Helmholtzzentrum f\"ur Schwerionenforschung GmbH, Institute of Particle and Nuclear Studies, KEK, and J-PARC Center. 
The authors gratefully acknowledge the support of the CBM Collaboration and the FAIR Phase-0 program at GSI, which enabled the experimental activities and infrastructure used in this work.
We gratefully acknowledge the support of the Detector Laboratory at GSI  for the production of the E16-STS detector modules.
We gratefully acknowledge the support and assistance of all personnel involved in the PF-AR Test Beam Line of the Instrumentation Technology Development Center at KEK.
This work was supported in part by 
the European Union’s Horizon 2020 research and innovation program EURIZON,
the Bundesministerium f\"ur Forschung, Technologie und Raumfahrt (BMFTR, Germany), 
the GSI Helmholtzzentrum f\"ur Schwerionenforschung GmbH (Germany),
JSPS KAKENHI Grant Numbers JP20H05647, JP23H05440, JP21H01102 and 24H00236 (Japan), 
the Ministry of Higher Education and Science (Poland).

\appendix

\bibliographystyle{unsrtnat} 
\bibliography{example}

\end{document}